\documentclass{article}

\usepackage{arxiv}

\usepackage[utf8]{inputenc} 
\usepackage[T1]{fontenc}    
\usepackage{hyperref}       
\usepackage{url}            
\usepackage{booktabs}       
\usepackage{amsfonts}       
\usepackage{nicefrac}       
\usepackage{microtype}      
\usepackage{lipsum}		
\usepackage{graphicx}
\usepackage{natbib}
\usepackage{doi}

\usepackage{epstopdf, epsfig}
\usepackage{placeins}
\usepackage{dcolumn}
\usepackage{bm}
\usepackage[dvipsnames]{xcolor}
\usepackage{comment}
\usepackage{hyperref}
\usepackage{amsmath}

\title{Unjamming strongly compressed particle rafts}

\author{ {Gregor Plohl} \\
	\And
{Mathieu Jannet} \\
	\And
		{Carole Planchette}  \\
	\texttt{carole.planchette@tugraz.at} \\
	\\
	Institute of Fluid Mechanics and Heat Transfer, Graz University of Technology, A-8010 Graz, Austria}

\renewcommand{\shorttitle}{Unjamming strongly compressed particle rafts}

\hypersetup{
pdftitle={Unjamming strongly compressed particle rafts},
pdfsubject={Fluid mechanics, Granular Matter},
pdfauthor={Gregor Plohl, Mathieu Jannet and Carole Planchette},
pdfkeywords={particle-laden interfaces, capillary,  compression, relaxation, folds},
}

\begin{document}
\maketitle

\begin{abstract}

We experimentally study the unjamming dynamics of strongly compressed particle rafts confined between two fixed walls and two movable barriers. The  back barrier is made of an elastic band, whose deflection indicates the local stress. The front barrier is pierced by a gate, whose opening triggers local unjamming.  The rafts are compressed by moving only one of the two barriers in the vicinity of which folds form. Using high speed imaging, we follow the folded, jammed, and unjammed raft areas and measure the velocity fields inside and outside of the initially confined domain. Two very different behaviors develop. For rafts compressed  by the back  barrier, only partial unjamming occurs. At the end of the process, many folds remain and the back stress does not relax. The flow develops only along the compression axis and the particles passing the gate form a dense raft whose width is the gate width.  For rafts compressed at the front, quasi-total unjamming is observed. No folds persist and only minimal stress remains, if any. The particles flow  along the compression axis but also normally to it and form, after the gate, a rather circular and not dense assembly. We attribute this difference to the opposite orientation of the force chain network that builds up from the compressed side and branchs. 
For rafts compressed at the gate side, keystone particles are immediately removed which enhances local disentanglement and leads to large scale unjamming. In contrast, for back compressed rafts, the force chain network redirects the stress laterally forming arches around the gate and resulting in a limited unjamming process. 
\end{abstract}

\maketitle


\section{Introduction}


Capillary adsorbed particles  at liquid interfaces have been since long used in emulsions and foams \cite{Ramsden1903, Pickering1907, Pitois_2019}.  Their potential for further applications, especially for producing   bijels \cite{Binks2002,  Herzig_2007, Cates_2008}, chemical-free reversible encapsulation \cite{Aussillous2001, Abkarian_2013, Jambon_2018, Pike1211} and membranes \cite{Kralchevsky_2001} has renewed interest in them. In this context, one key-attribute of the particles is their capacity to stabilize interfaces. This stabilization partly results from  the large reduction of surface energy caused by the  particle adsorption.  Further stabilization originates from the physical barrier the particles build that prevents direct contact with other surfaces. 
The effects resulting from  these two stabilization mechanisms  remain  challenging to characterize at the macroscopic scale. The most common way consists in  characterizing these effects via the description of  the interfacial mechanical properties. Thanks to intensive research, these properties are fairly well outlined for quasi-static regimes and moderate compression. Yet, for general conditions, they remain  poorly understood and therefore poorly predictable. This lack of knowledge is particularly profound for large strains and large strain rates or when strong gradients of particle density exists. For these conditions, commonly encountered in natural situations and industrial processes \cite{Garbin_2019},  barely no data exist. Consequently, key aspects such as self-healing capacity, self-healing dynamic and  processability of these interfaces remain widely  unexplored. Our paper aims to shed some light onto these topics.

Before detailing our method and findings, it is helpful to recall the existing knowledge  about the mechanical properties of particle-laden interfaces. 
At moderate particle density, the interface is viscous. It becomes viscoelastic upon the particle network percolation, and finally behaves as a solid for larger particle density \cite{reynaert2007interfacial, Cicuta2003, Lagubeau2010}. The transition  toward a solid-like behavior is associated to a pressure collapse \cite{Aveyard2000, monteux2007determining} and attributed to a jamming process \cite{liu1998jamming}, a phenomenon common to other athermal assemblies, also called soft glassy materials \cite{sollich1997rheology, hebraud1998mode}. If further compressed, the interface buckles giving rise to regular wrinkles whose amplitude regularly increases, until  one of them dramatically grows and eventually collapses into a large fold.

To date,  most studies were motivated by  practical interest such as bubble dissolution arrest and foam stabilization \cite{beltramo2017arresting, taccoen2016probing, Abkarian2007, timounay2017viscosity} and therefore focused on the  mechanical properties of  interfaces  prepared between the jamming and folding transitions.  In this range, it has been shown that the interface can be modelled as a continuous media. This approach is based on the  buckling of the interface subjected to quasi-static uni-axial compression. Similar buckling is observed for monolayers of irreversibly adsorbed molecules \cite{milner1989buckling},  membranes \cite{helfrich1973elastic} and solid sheets \cite{Cerda2003, Pocivavsek2008a}.
The energy minimization selects a wavelength which results from the competition between bending and gravitational ironing.  The selected wavelength gives the elastic Young modulus (for 3D approach) or the elastic bending modulus (for 2D approach) of the particle-laden interface \cite{Vella2004}. 
This elastic modeling proved to be valid for dynamic regimes too \cite{Planchette2012a} but fails when mixtures of small and large particles were used. In this case,  deviations from the expected modulus are observed which were attributed to possible variations of the stress transmission efficiency, potentially caused by the coexistence of different types of particle-particle contacts \cite{pre_bidisperse}. This observation with others, such as: finite size effects in uni-axial compression \cite{Cicuta2009}, mechanical response under compression \cite{Pitois2015, taccoen2016probing},  raft fractures \cite{Vella2006}, inhomogeneous stress propagation \cite{basavaraj2006packing, saavedra2018progressive}, plasticity \cite{Jambon_2018}… evidence the limit of the elastic model and points toward the granular character of these interfaces.
Indeed, these limits are probably related to the framework successfully developed for 2D granular media. In  dry granular matter, “force chains” emerge which have been experimentally observed and numerically computed \cite{tordesillas2011, peters2005}. These force chains are  suspected to build up  in particle rafts too, where they could significantly modify  the dynamical response  \cite{pre_bidisperse, planchette2018rupture}. Yet, dedicated investigations remain scarce and the present paper constitutes a first step in this direction.

More precisely, this work aims to understand how the above mentioned  granular character of the interface,  evidenced for moderate strains and mostly under quasi-static regime, affects its mechanical  properties  when subjected to  large strains or fast strain rates.
Said differently,  is the granular character of the interface relevant for the relaxation of strongly compressed rafts? and what could be its effects?
While answering these questions undoubtedly leads to a gain of  fundamental knowledge, it also addresses a crucial practical aspect, known under the term of "self-healing".   In contrast to commonly used surfactants, the particles are irreversible absorbed to the interface and can therefore not constitute  reservoirs in the bulk from where to migrate in order to feed freshly created interface areas  \cite{binks_2002, garbin_2013}. While this aspect is positive to arrest bubble dissolution, it remains challenging for expending interfaces.  A good strategy to efficiently overcome this challenge  might be to  establish particle reservoirs outside the bulk, and more particularly in  interfacial folds.  In this context, the questions can be reformulated as such: can   particles located within folds become available to stabilize uncovered areas,  providing self-healing capacity to these interfaces?  Is this capacity, i.e. the portion of stored particles that can be efficiently released, influenced by the granular character of the interface? What about  the kinetic of this supply, i.e. the self-healing dynamic? 

We  answer these questions by studying the dynamical  relaxation of strongly compressed particle-laden interfaces and show that two typical behaviors are obtained, which are solely selected by the direction of compression. First, 
 we  describe  the experimental methods. The results    - obtained  for two square rafts of same compression  performed either from the back or the front side - are then detailed. 
We show that similar findings are also obtained for  less compressed rafts. We attribute the different behaviors to a purely granular effect, namely the orientation of the chain force network, which either favors unjamming via the removal of keystone particles, or prevents it by redirecting the stress toward the sides.  The paper ends with the conclusions which underline the consequences of our findings in terms of  interface processability, self-healing capacity, and dynamic.

\section{Experimental methods}


\subsection{Set-up}

\begin{figure}[]
\centering
\includegraphics[width=8cm]{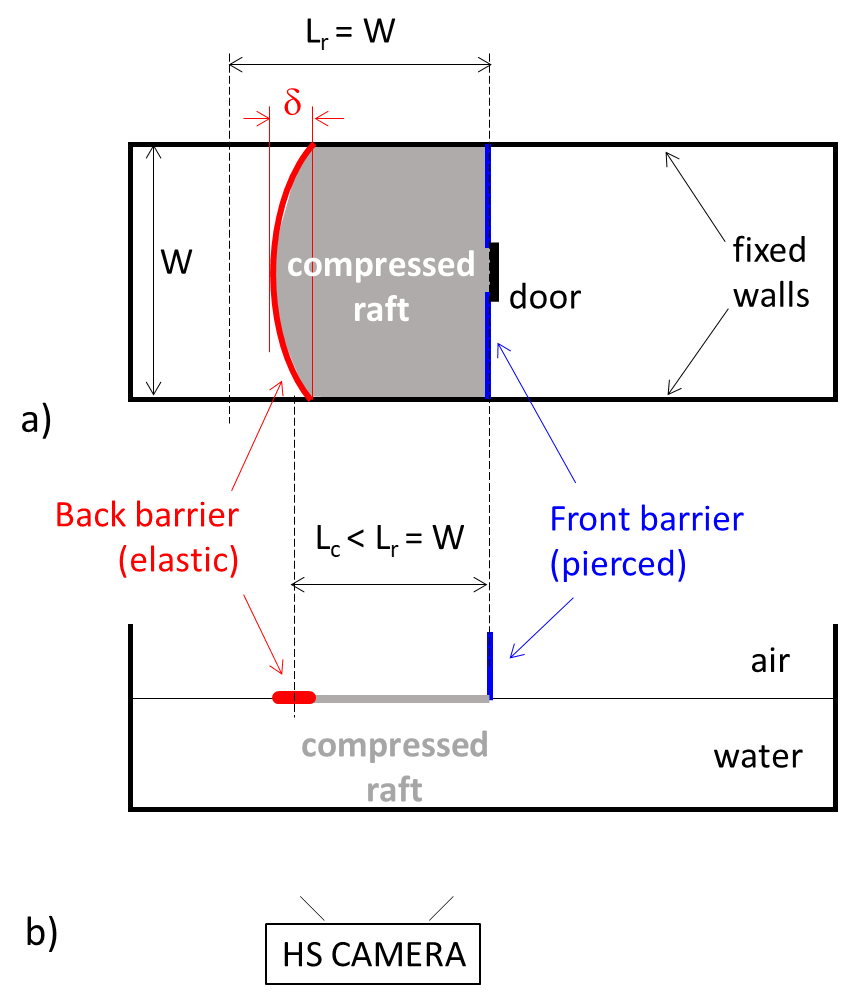}
\caption{Experimental set-up sketched from  (a)  top and (b) side view. The particle raft relaxed length, $L_r$, is equal to its width $W$. The raft is compressed by moving either the back barrier (red) or the front one (blue) to obtain a length $L_c < L_r$. The elastic deflection $\delta$ (exaggerated here) enables to calculate the local stress. }
\label{fig:1}
\end{figure}

As sketched in Fig. \ref{fig:1}, the experimental set-up consists of a rectangular trough with two fixed parallel walls separated by 6 cm and two movable barriers that can be translated along the two walls to compress the confined particle raft. 
The so-called "back" barrier is made of  an elastic string placed in the plane of the interface, perpendicularly to the side walls. The  string is produced in house by injecting a (1:1) mixture of  Elite Double 8 basis and catalyst (Zhermack Spa)
into a glass capillary, which is manually removed after the elastomer reticulation has been completed. The elastic is then fixed to a 6 cm broad structure and calibrated using known weights to obtain its Young modulus, for details please see Appendix. During the experiments, the elastic deflection, $\delta$, is measured and used to calculate the stress developing  at the back side of the raft, see Appendix. The second barrier, refereed to as "front" barrier, is pierced in its center by an  orifice of width $w=1cm$. A vertically sliding gate made of a thin plate enables to maintain the orifice closed. By suddenly lifting the gate,  the orifice gets unblocked and  the applied stress is locally released. 

\subsection{Particles}

The particles are sieved and silanized glass beads. The diameter distribution measured over pictures of more than 1000 particles corresponds to a Gaussian with a mean of  107 $\mu m$ and a standard deviation of 8.4\%. After having cleaned the particles with a piranha mixture,  silanization is performed  using solution of trichloro-perfluorooctylsilane in anhydrous hexane. All chemicals were purchased from Sigma-Aldrich and used as received. The resulting contact angle measured on single particles placed at the apex of a pendant drop is found to be $107^{\circ} \pm 10^{\circ}$.

\subsection{Experimental procedure}

The first step of the experimental procedure consists in obtaining square rafts. To do so, the trough is filled with distilled water and particles are sprinkled on the interface between the barriers. By gently blowing on them, we insure that they distribute in  a monolayer. This monolayer is then compressed and decompressed by moving any of the two barriers and the quantity of particles is adjusted in order to have a square relaxed raft.   We define the raft as relaxed as soon as the stress  measured  during decompression vanishes. Noting $L_r$ the relaxed length and $W$ the trough width, we have $L_r=W=6 cm$. Once a square raft is formed, the distance between the barriers is increased and the raft is annealed by stirring the particle assembly. The raft is then compressed again, but  \textbf{by moving only one of the two barriers}, the other one remaining fixed until the end of the experiment. 
The state of compression is then given by $K=(L_r-L_c)/L_r$ with  $L_c$ the distance between the two barriers in their final position. Finally,  the relaxation is triggered by suddenly opening the gate. A high speed camera placed under the raft records its evolution. Practically, 3000 frame per second are recorded. The images have a resolution of 14.1 $\mu m/pixel$. The movies are then analysed to provide different measurements.

\begin{figure*}[ht]
\centering
\includegraphics[width=0.95\textwidth]{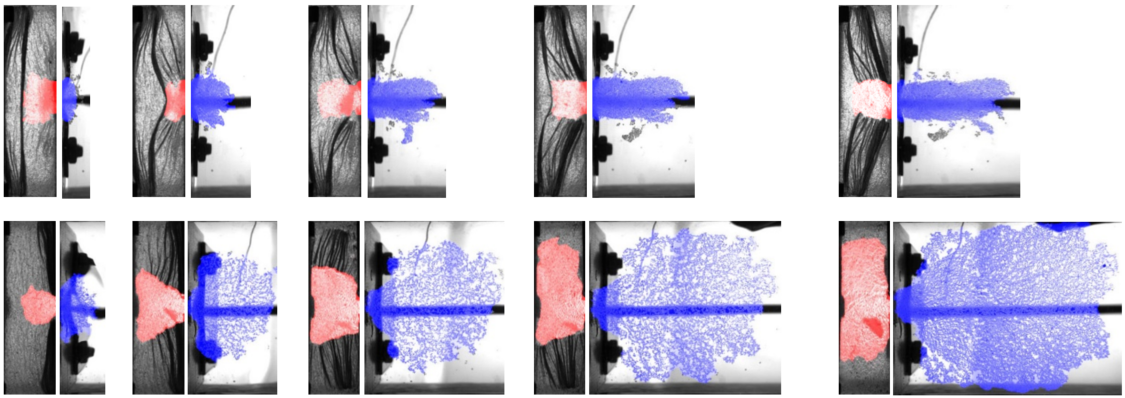}
\caption{Image sequences of strongly  compressed rafts  ($K=67\%$) after local release of the stress achieved by lifting the gate of the front barrier. The opening instant is taken as time origin. The red color indicates unjammed particles area while the blue shows the "escaped" particles.   Top: back compression (subscript b) with from left to right: $t_{b,1}=70$ ms, $t_{b,2}=120$ ms, $t_{b,3}=250$ ms, $t_{b,4}=360$ ms and $t_{b,5}=430$ ms. Bottom: front compression (subscript f) with from left to right: $t_{f,1}=80$ ms, $t_{f,2}=410$ ms, $t_{f,3}=630$ ms, $t_{f,4}=1600$ ms, $t_{f,5}=2800$ ms.}
\label{fig:2}
\end{figure*}

\section{Measurements}

\subsection{Unjammed, jammed, folded and escaped areas}\label{sec:masks}

During the raft relaxation, several quantities are measured, which first need to be defined. The illustrative image sequences displayed in Fig. \ref{fig:2} are useful to do so. Let us  first consider what happens in the initially confined domain. After the gate opens,  some particles locally unjam. The corresponding surface area, $A_{uj}^*$,  colored in red in Fig. \ref{fig:2}, is tracked using ImageJ and the machine learning plug-gin called Trainable Weka Segmentation \cite{arganda2017trainable}. After normalisation by the initial confined domain area $A_c=W L_c$, we obtain $A_{uj}=A_{uj}^*/A_c$. Per definition, the particles that remain jammed occupy the normalized area $A_{j}= 1-A_{uj}$. They can take the form of macroscopic folds or remain in the plane of the interface. To characterize this distribution, we measure the so-called folded surface area $A_{f}^*$.  Practically, the folds correspond to interface portions which make a significant angle to the horizontal plane and therefore appear dark on the back-lighted pictures. Thus, $A_{f}^*$ is obtained by using a threshold function and is then normalized by $A_c$ to provide $A_{f}$.  Note that this term refers to the projection of the folded surface and not to the surface area contained in these folds, which is not accessible with our images.

Let us now look to what happens outside of the initially confined domain. Some of the unjammed particles flow trough the orifice and migrate further to form a more or less dense assembly, colored in blue in  Fig. \ref{fig:2}. The surface area occupied by this assembly $A_{e}^*$, is tracked using a threshold function applied after subtraction of the background taken in the absence of particles.   The normalized escaped area $A_{e}$ is obtained by dividing $A_{e}^*$ with $(L_r-L_c)W=A_r-A_c$, i.e. with the difference between the relaxed raft area and the one of the confined domain. This difference represents the surface that  the excessive particles would occupy if forming a dense relaxed raft. 

\subsection{Back stress}

Additionally,  the deflection of the elastic barrier $\delta$, is recorded and used to compute $\Pi$, the lineic pressure (or stress) developing at the back of the raft. For details about the conversion of $\delta$ into $\Pi$, please read the Appendix.

\subsection{Velocity fields}\label{sec:piv}
Finally, we use the PIVlab routine of Matlab \cite{thielicke2014pivlab, 
thielicke2021particle} to gain information about the velocity fields developing in these systems. In practise, we perform three types of Particle Image Velocimetry (PIV). 

The first one focuses on the flow inside the confined area, and more precisely within the unjammed area. To avoid wrong interpolations with movements  taking place in the jammed raft, such as fold translation, masks corresponding to the identified unjammed areas  (red regions in Fig. \ref{fig:2}) are applied, limiting the analysis to the region of interest.
The second analysis is  similar and consists in performing PIV  outside of the confined domain using masks that correspond to  the escaped assembly (blue regions in Fig. \ref{fig:2}. Note that we further reduce the masks of about 5 mm on the orifice side, since the nonuniform background of this zone causes errors in the PIV. 

These two PIVs, {typically performed with a time resolution of $300 Hz$, }
provide - for every instant - the velocity field inside and outside the confined domain. Due to the large number of frames, the instantaneous velocity fields are difficult to visualize and interpret. To facilitate their analysis, we divide each relaxation in four sub-phases, and compute the  distribution of the velocity components parallel and normal to the compression axis ($v_{\parallel}$ and $v_{\perp}$) for each sub-phase. The  sub-phases  correspond to the time lapse between two consecutive images of Fig. \ref{fig:2}, i.e  to  $t_{b, i} \leq T_{b,i} \leq t_{b, i+1}$ and  $t_{f, i}  \leq T_{f,i} \leq t_{f, i+1}$, with $1 	\leq i \leq 4$,  for the back and front compression, respectively.

The third analysis deals with the velocity profile at the orifice. To obtain this information, we limit the analysis to a small zone covering the orifice and increase the time resolution  to at least $600 Hz$. The width of the analysed area is the orifice width and its length is fixed to $10mm$. For each picture, the obtained velocity field is then projected along the compression direction to provide the velocity profile $u(x)$, with $-w/2 \leq x \leq  w/2$. The flow-rate equivalent mean velocity $\overline{u}$ is then calculated as $ \overline{u}= \frac{1}{w}\int_{-w/2}^{w/2}{u(x)dx}$.

\section{Results and discussion}

Here and in the rest of this article, results obtained with front compression are presented as solid lines and full symbols while those corresponding to back compression are represented by dashed lines and empty symbols. The red color systematically indicates quantities related to the confined domain while blue is used for quantities defined outside.

\subsection{Detailed analysis of rafts compressed at 67\%}

In this section,  we present the results obtained for two rafts compressed at 67\% by moving solely   the front barrier or solely the back barrier. 
Illustrative image sequences are reproduced in Fig. \ref{fig:2}.  To facilitate the observation,  the unjammed area found in the confined domain, $A_{uj}^*$,  is systematically colored in red and  the surface occupied by the "escaped" particle assembly,  $A_{e}^*$, in blue. These colored masks, automatically obtained by the image treatment (section \ref{sec:masks}), are in very good agreement with the areas detected by eyes and can therefore be used for further analysis. The two rafts behave differently. For the back compression (top sequence), only a small zone found immediately behind the orifice  umjams. The unjammed particles flow through the orifice, along the compression axis, to form a dense assembly whose width is the orifice width. The relaxation process stops while many folds are still visible. In contrast, the relaxation of the raft compressed from the front (bottom sequence) leads to important unjamming, which extends to almost all the confined domain. The escaped particles  form a large assembly of reduced density and broad width. No folds remain indicating a full relaxation of the raft. As the caption timestamps indicate, the relaxation of the back compressed raft is much shorter. It almost stops after  $0.5s$, i.e. long before the one of the front compressed raft finishes, after more than $3s$.

\begin{figure}[h]
\centering
\includegraphics[width=0.8\linewidth]{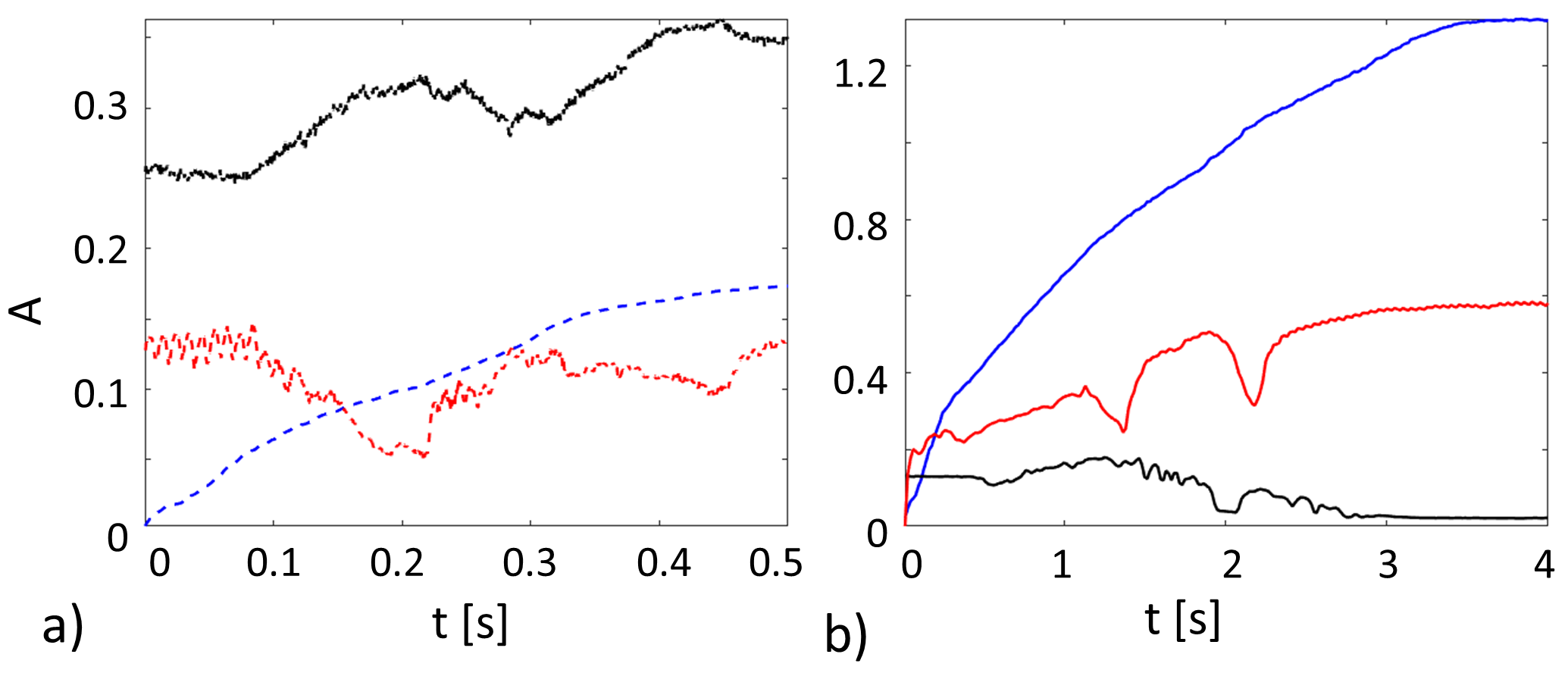}
\caption{ Temporal evolution of $A_{uj}$ (red), $A_{f}$ (black)  and $A_{e}$ (blue) for (a) back- and   (b) front- compressed raft.}
\label{fig:areas}
\end{figure}

To quantify this observation, we plot the temporal  evolution of the normalized  unjammed area $A_{uj}$  (red),  folded area $A_{f}$  (black) and escaped assembly area $A_{e}$  (blue). The curves, shown in Fig. \ref{fig:areas}, confirm the qualitative findings. For the raft compressed from the back, slightly less than 15\% of the initially confined area  unjams and approximately 35\% of the confined area appears folded at the end of the process. Interestingly, the unjammed area does not significantly grow but rather fluctuates around its mean value. These fluctuations originate the folds dynamics. The folds,  initially found at the back (Fig. \ref{fig:2}, $t_{b,1}=70ms $) migrate  towards the orifice and transiently reduce the unjammed area ($t_{b,2}=120ms $) before  disintegrating ($t_{b,3}=250ms $),  explaining likewise the correlation between the local minimums of $A_{uj}$, at $\approx 0.2s$ and $\approx 0.45s$, and the local maximums of $A_{f}$. 
Finally, the particles that escape covers between 15\% and 20\% of the surface excessive particles would cover if forming a dense relaxed raft. This means that the large majority  of the particles stored in the folds (at least 80\%) cannot be made available to supply the neighboring surface initially free of particles. From a practical point of view, the self-healing capacity of such interfaces appear very limited under the present conditions. 

Let us now consider the  raft  compressed to the same level but from the front side. The  evolution of the unjammed, folded and escaped areas appear to be totally different, see  Fig. \ref{fig:areas}(b).   About 60\% of the initially confined area  unjams. The folded area, which represents at maximum 15\% of the confined area totally disappears. Here as well, the evolution of these two quantities seem to be correlated as indicated by the coincidence of  the local minimums of $A_{uj}$  with the local maximums of $A_{f}$, found around 1.4s and 2.2s. Yet, in contrast to back compressed raft, these events do not correspond to the migration, expansion and disintegration  of individual folds but to the elimination of two larger folded blocks, found on both sides of the orifice. The first elimination takes place between the third ($t_{f,3}=0.63s$) and fourth ($t_{f,4}=1.6s$) pictures of Fig. \ref{fig:2}  while the second one occurs between the fourth ($t_{f,4}=1.6s$) and fifth ($t_{f,5}=2.8s$) pictures. Beside these two bumps, $A_{uj}$ continuously increases while $A_{f}$ decreases.  Almost all  particles initially stored in folds unjam and migrate trough the orifice to cover a surface that is 1.3 times the excessive surface given by $A_r-A_c$. While a value larger than 1 can first be surprising, it is well explained by the fact that the assembly of escaped particles is less dense than the relaxed raft. The relaxed raft density  is close to the one of jamming,  defined as $\varphi_{j}=\pi / 2\sqrt{3} \approx0.91$. The one of the escaped assembly can be only roughly estimated from our pictures, providing $\varphi_{e} \approx 0.74 $, in agreement with  0.76, the value required to obtain "particle surface conservation" under total unjamming and given by  $\varphi_{e} = \varphi_{j} /A_{e}$. 
These results clearly show that if the compression direction is favorable, the self-healing capacity of particle-laden interfaces can become very important and approach its theoretical maximum. This maximum is found when all particles initially stored into folds are efficiently released and mobilized to cover initially "free" interface.

At this stage, it is difficult to draw conclusions about the self-healing dynamics. For front compressed raft, only $0.12s$ are required to obtain $A_{e}=15\%$ by comparison to $0.33s$  for back compressed raft. Yet, this rate quickly slows down. If calculated over a longer period of time ($2s$), it is found to be $\dot{A}_{e}=0.5s^{-1}$,  comparable  to  $0.45s^{-1}$, the rather constant rate obtained for back compression. 

\begin{figure}[h]
\centering
\includegraphics[width=0.8\linewidth]{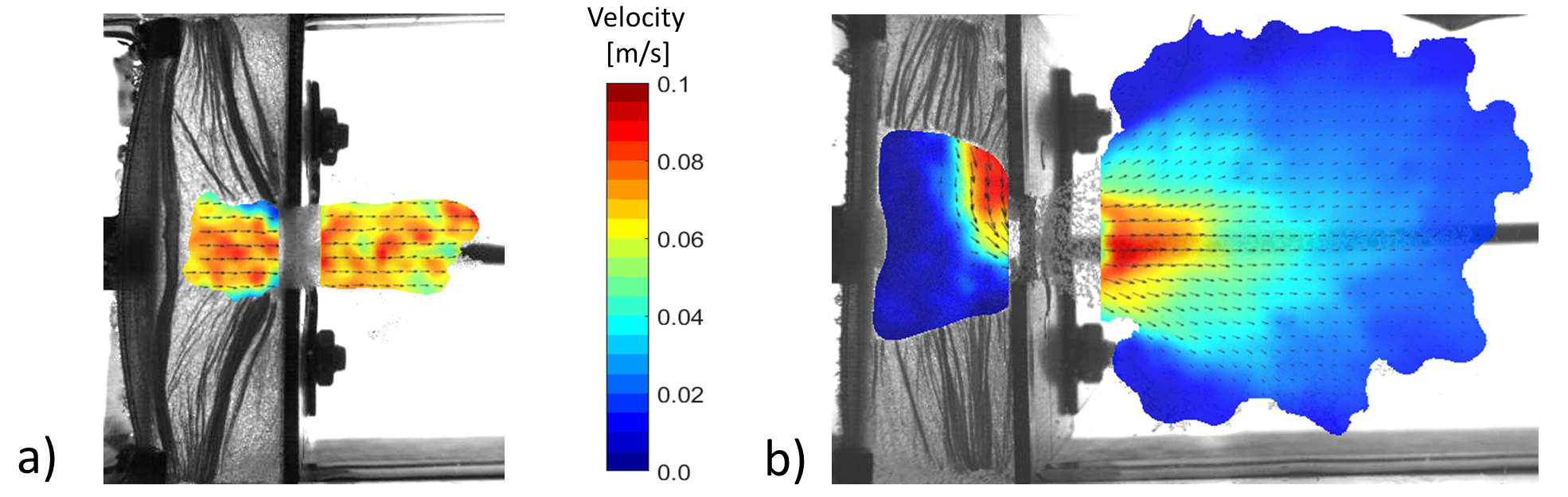}
\caption{Exemplary results of the PIV for (a) back compressed raft, $t_b=0.28s$ and (b) front compressed raft, $t_f=1.28s$. }
\label{fig:PIV_1}
\end{figure}

To better understand the relaxation dynamics, we perform PIV on the unjammed areas found  inside and outside  the initially confined domain. Representative snapshots of the results are shown in Fig. \ref{fig:PIV_1}. For the back compressed raft, both the velocity magnitude and direction are uniform. The vectors clearly indicate a flow in the compression direction, from the back of the raft to the orifice and beyond. In contrast, the front compression gives rise to local acceleration, especially close to the orifice, and components perpendicular to the compression direction develop. 
Interestingly, the velocity field of the unjamming particles inside the confined domain is not symmetric. Indeed, the instant chosen here ($1.28s$) corresponds to the beginning of the already mentioned elimination of one of the two  folded blocks found on each side of the orifice (here, up on the picture).   

\begin{figure}[]
\centering
\includegraphics[width=0.4\linewidth]{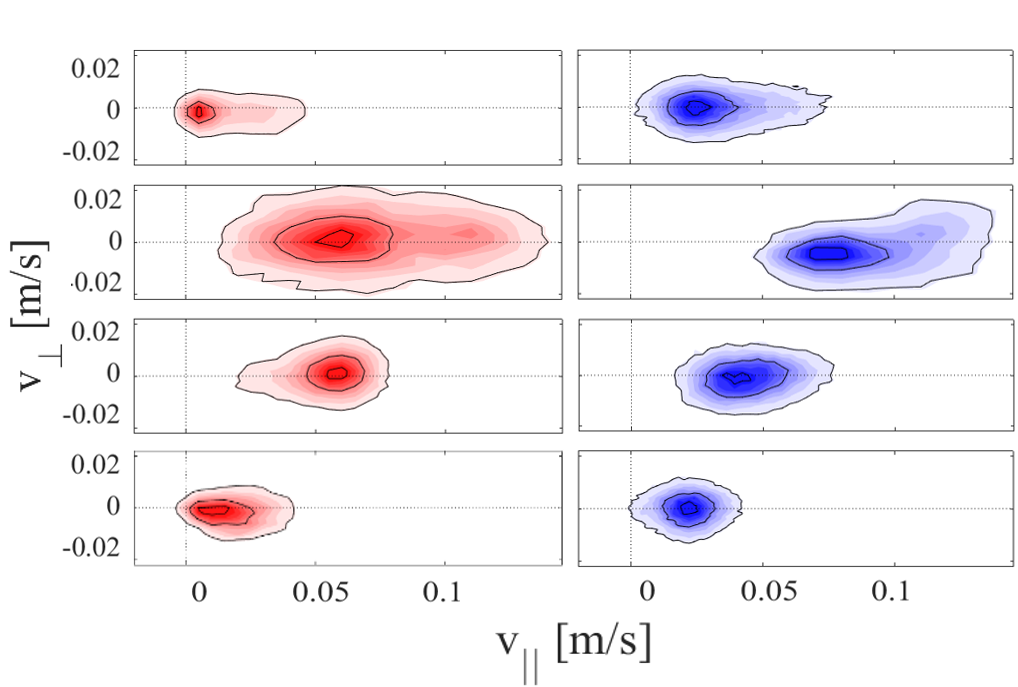}
\includegraphics[width=0.4\linewidth]{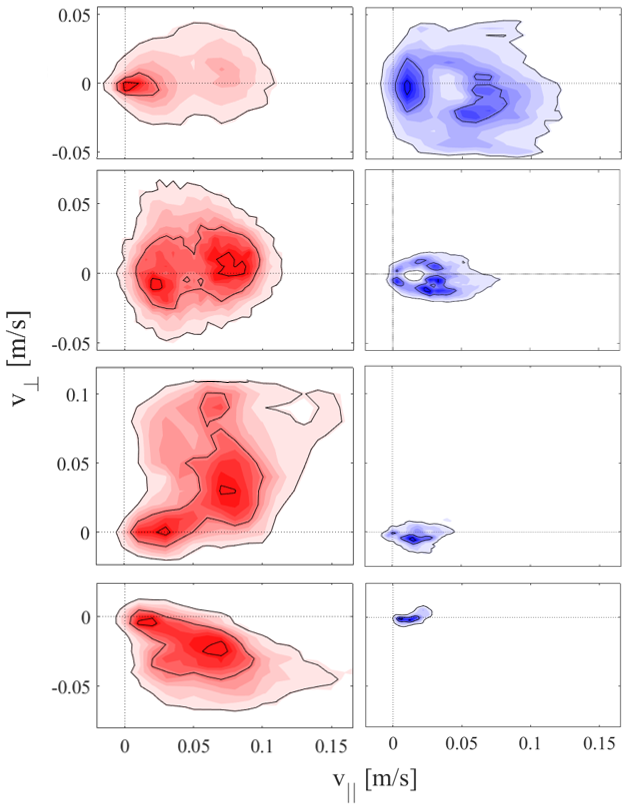}
\caption{Normalized velocity distributions  for left: back compressed raft, and rigth: front compressed raft. Red: unjammed area inside the confined domain; blue: escaped particles outside of it. From top to bottom, the sub-phases are: $T_{b,1}$, $T_{b,2}$ ,$T_{b,3}$, $T_{b,4}$  for the back compressed raft; and $T_{f,1}$, $T_{f,2}$ ,$T_{f,3}$, $T_{f,4}$ for the front compressed raft. The contours correspond to 0.1, 0.5 and 0.9 of the maximum value.}
\label{fig:map_1}
\end{figure}

To visualize the PIV results obtained over the whole process, it may be useful to compare the velocity distribution obtained on the  sub-phases  defined in section \ref{sec:piv}. They are shown in Fig. \ref{fig:map_1}, left and right,  for back and front compressed raft, respectively. For the back compressed raft, the velocity fields inside (red) and outside (blue)  of the confined domain are always very similar, confirming that the unjammed particles form a cohesive assembly, which moves as a block. The velocity component perpendicular to the compression axis, $v_{\perp}$, is centered in zero and shows very small fluctuations, the 50\% contour being comprised in $\pm 0.01m/s$. This is not the case of the parallel component, $v_{\parallel}$, which has positive values  along the whole process. The flow from the back to the front shows some variations with a maximum  during $T_{b,2}$, i.e. when the first fold disintegrates. Its mean value ($0.07m/s$) and the fluctuations around it ($-0.03m/s$ and $+0.07m/s$) then decrease until the end of the process.

The velocity fields that develop for the front compressed raft are very different, see Fig. \ref{fig:map_1}.  The  correspondence between inside and outside is lost, which may indicate a certain independence of the two flows. Inside the confined domain (red),  significant movements develop perpendicular to the compression axis leading to  $v_{\perp}$ values as large as $v_{\parallel}$ values. The flow is not symmetric across the compression axis, see for example  $T_{f,3}$ or $T_{f,4}$ for which $- 0.01 < v_{\perp} < + 0.11 m/s$ and $-0.06 < v_{\perp} < 0.01 m/s$, respectively. This profound asymmetry corresponds to the successive dislocation of two large folded blocks found on each side of the orifice. While passing through the orifice, the particles    recover a symmetric velocity field (see blue maps). One can suppose that the shear they experience at this point overcome the  capillary attraction, forcing them to rearrange and explaining the reduced density. Finally,  the focus of the velocities toward zero  can rather be attributed to  a normalizing effect than to changes in the flow close to the orifice. The increasing investigated surface area  ($A_e$) contains an increasing number of points with low velocity, typically located at the assembly periphery, and therefore focuses the normalized distribution.

\begin{figure}[h]
\centering
\includegraphics[width=0.4\linewidth]{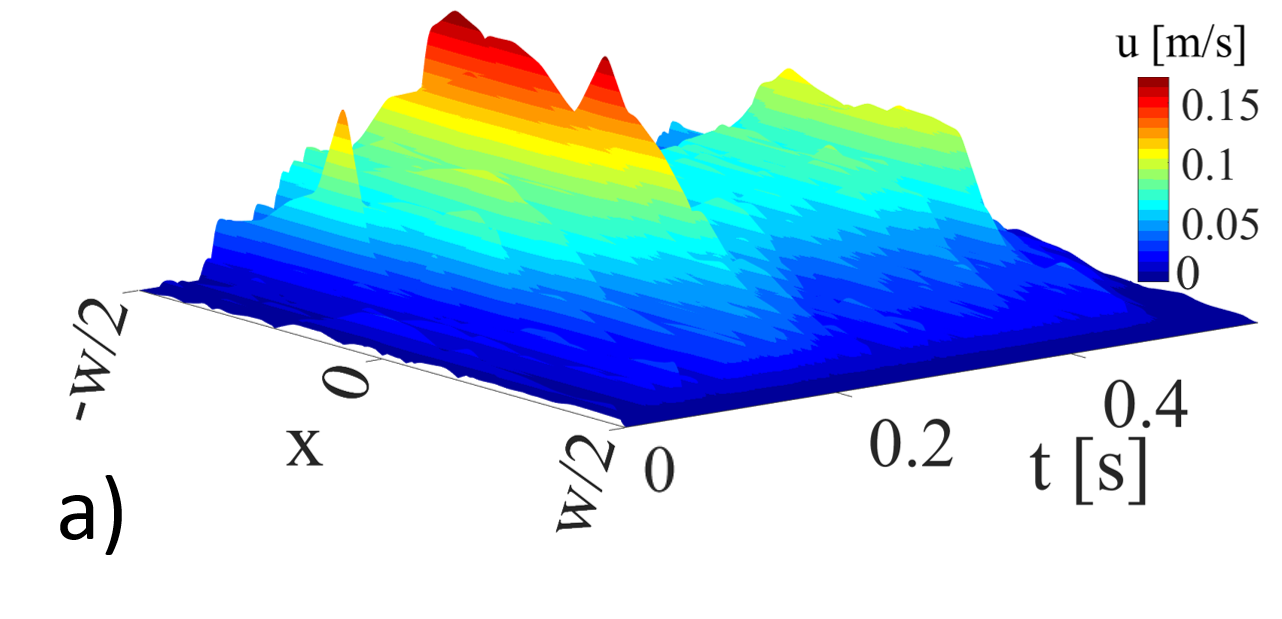}
\includegraphics[width=0.4\linewidth]{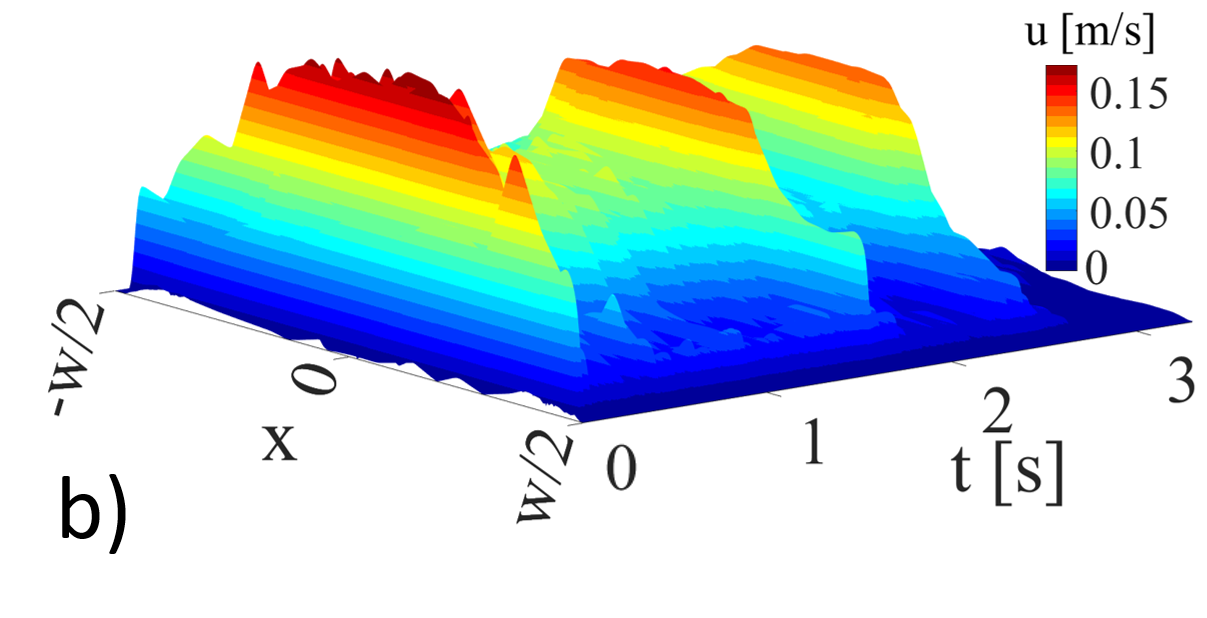}
\caption{Velocity profiles at the gate, $u(x)$, as a function of time, $t$,  for (a) back and (b) front compression.}
\label{fig-GateVelProfiles3d}
\end{figure}

Despite the important disparities of the velocity fields of the front and back compressed rafts, the velocity profiles at the orifice seem first rather similar. The profiles displayed in Fig. \ref{fig-GateVelProfiles3d} do not evidence obvious differences. The shape of the profile, the value of the maximum velocity and the importance of the fluctuations are comparable in both cases.

\begin{figure}[ht]
\centering
\includegraphics[width=0.45\linewidth]{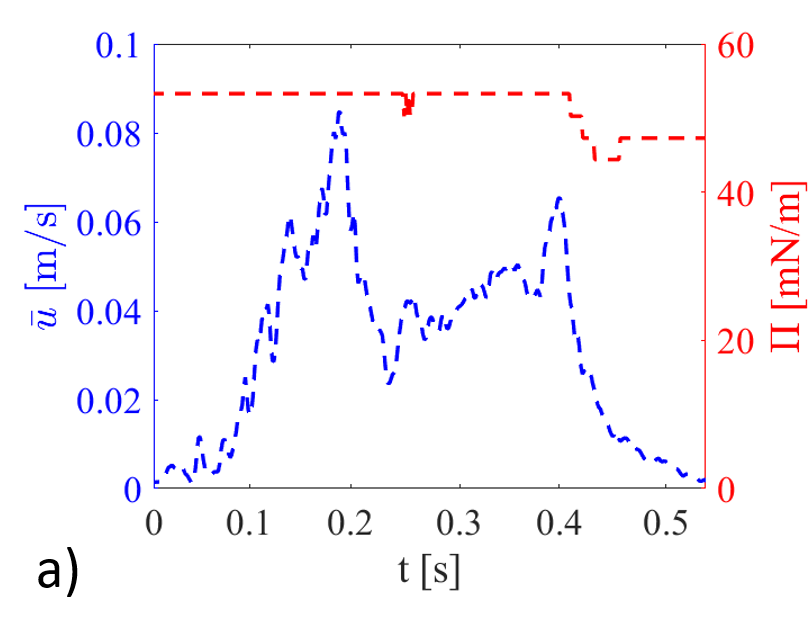}
\hfill
\includegraphics[width=0.45\linewidth]{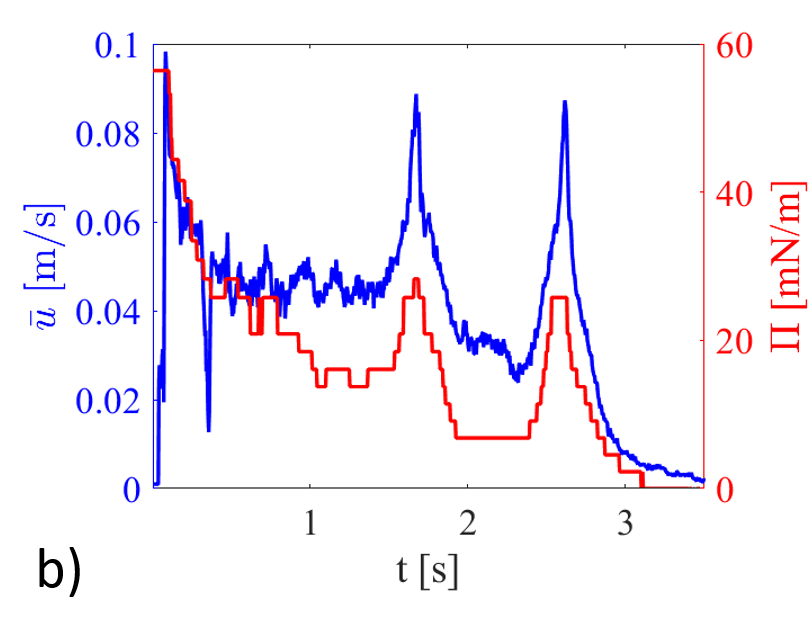}
\caption{Temporal evolution of $\overline{u}$ (blue) and $\Pi$  (red) for (a) back compression and (b) front compression. }
\label{fig-UandRubber1}
\end{figure}

 To better analyse the particle flux and its fluctuations, we now focus on  the mean velocity $\overline{u}$. The results are plotted in Fig. \ref{fig-UandRubber1} together with $\Pi$, the stress measured at the raft back. Two points are worth being discussed. First, the different evolution of $\overline{u}$ for the front and back compressed rafts. For the back compression, after a short transitory phase and before the relaxation stops,  $\overline{u}$ seems rather constant. The two peaks, attributed to the disintegration of  two successive folds (see Fig. \ref{fig:2}-\ref{fig:areas}), can be identified but no clear increase or decrease of $\overline{u}$ over time can be seen. It is not the case for the front compressed raft for which $\overline{u}$ clearly decreases over time. Fluctuations are perturbing this evolution but they can be well explained by the already mentioned elimination of the two folded blocks found on each side of the orifice.
 This confirms that the compression side does not only influence the self-healing capacity but also its kinetics. The second important point evidenced by Fig. \ref{fig-UandRubber1} concerns the stress measured at the back of the raft. The latter remains almost unchanged for the back compressed raft, indicating that an important fraction of the initially applied stress remains. This remaining stress  most likely originates the still compressed portions found on each side of the central unjammed corridor. A closer look at the curve indicates a slight and probably step-wise  decrease of $\Pi$ after each fold  disintegration. If confirmed, such variations are typical of a solid-like behavior, for which the stress and strain are linearly connected via the Young modulus.    For the front compressed raft, the stress totally relaxes. Interestingly, fluctuations are observed that correspond to the ones of $\overline{u}$, indicating a direct and almost instantaneous  transmission of the stress trough the entire raft. This could  be interpreted as a liquid-like behavior, for which the stress and strain rate are proportional, the proportionality coefficient being the viscosity.


\subsection{Generalisation and interpretation}\label{sec:general}

The detailed results presented above were obtained on two similar rafts and generalizing these findings requires more data. Thus,  these experiments were repeated with two other pairs of raft, compressed either from the front or from the back, with a compression level of 33\% and 50\%, respectively. 

\begin{figure}[ht]
\centering
\includegraphics[width=0.8\linewidth]{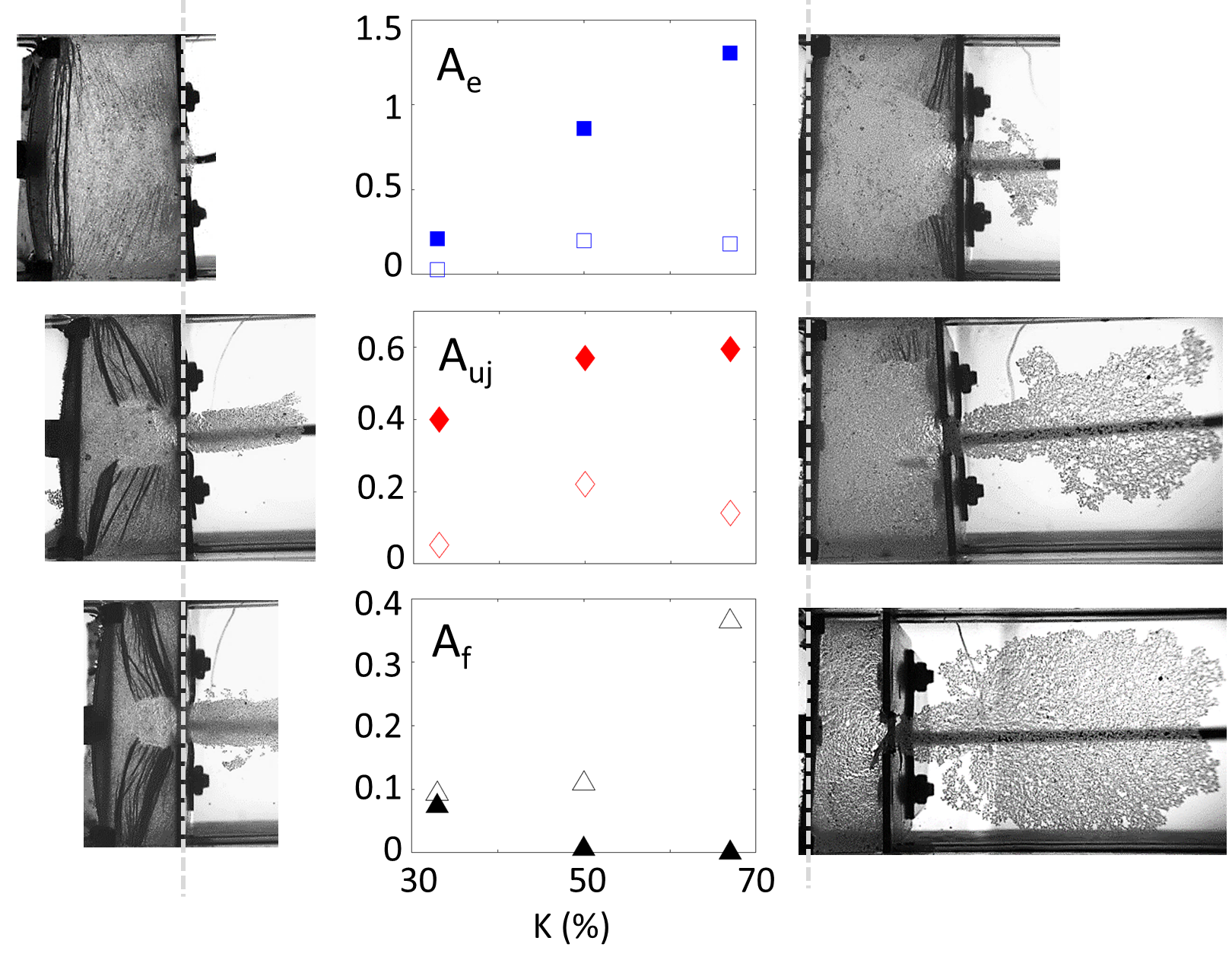}
\caption{ Plots (central raw): final values of $A_{e}$, $A_{uj}$ and $A_{f}$ as a function of $K$ with empty (full) symbols  for back (front) compression. Pictures:  final relaxation states obtained for back (left raw) and front (rigth raw) compressed rafts. From top to bottom, the raft compression is : K=33\%, 50\% and 67\%. }
\label{fig:last}
\end{figure}

The final relaxation states are characterized in Fig. \ref{fig:last} and appear to be  mostly fixed by the compression side. For rafts compressed on the back side (left picture raw), the relaxation is partial, folds remain  and the escaped particles form a rather dense assembly, whose width is the orifice width.  Similar rafts compressed from the front side give rise to a quite different outcome, see right picture raw. The relaxation is almost total, only a few folds remain if any, and the escaped particles form less dense assemblies whose width is much greater than the orifice width. A more quantitative description is shown in the central plots where $A_{e}$ (squares), $A_{uj}$ (diamonds) and $A_{f}$ (triangles) are reported as a function of $K$. Whatever the level of compression, the relaxation of the front compressed rafts (full symbols) is always much more complete than the one of the back compressed rafts (empty symbols). In other words, $A_{e}$ and $A_{uj}$ are systematically greater for front compression (factor 4 to 8 and 3 to 8, respectively)  while the opposite is observed for $A_{f}$. Focusing on the back compressed rafts, $K$ has a limited influence on $A_{e}$ and $A_{uj}$ but strong effects on $A_{f}$.
For front compression, the influence of $K$ on $A_{e}$ and $A_{uj}$ is significant. The greater the compression, the greater these two  quantities. The effect is stronger for $A_{e}$ than $A_{uj}$, which  could be explained by variations of the assembly density.  Indeed, the surface occupied by a given  number of escaped particles  is  modulated by the packing density of the assembly, which possibly decreases with increasing compression (or flow velocity). Folds are very limited,  which is expected given the large scale unjamming.


From a practical point of view,  these results  indicate that for the studied geometry, the self-healing capacity of particle laden interfaces is  principally a function of the compression direction. For a given direction, the greater the compression, the greater the fraction of stored particles that can flow and cover  initially particle free regions. This released fraction is always much larger for front than for back compression. 

\begin{figure}[h]
\centering
\includegraphics[width=0.45\linewidth]{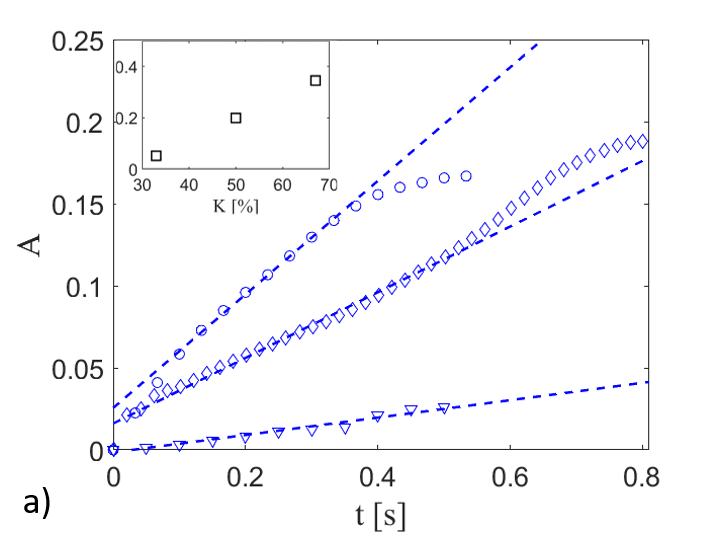}
\includegraphics[width=0.45\linewidth]{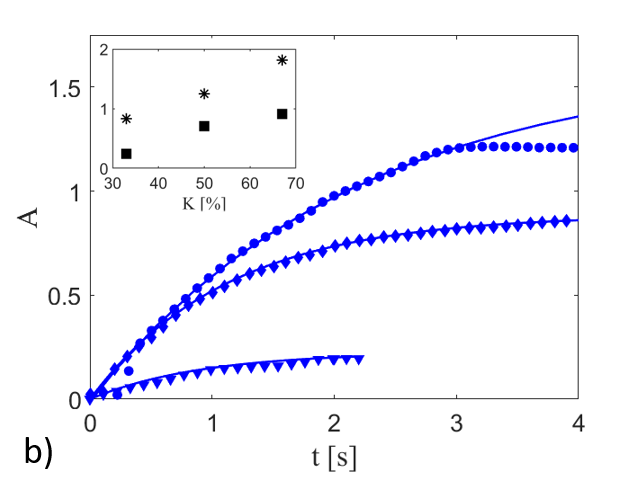}
\caption{  $A_{e} (t)$ for (a) back  and (b) front compressed rafts. Symbols are experimental data with triangles for $K=33\%$, diamonds for $K=50\%$ and circles for $K=67\%$. The dashed and continuous lines are linear and exponential fits given by Eq. (\ref{eq:discharge}) and (\ref{eq:discharge_2}), respectively. Insets: $\dot{A}_{e}(t=0)$ (squares) and $\tau$ (stars, if applicable) as a function of $K$. }
\label{fig:dynamics}
\end{figure}

To go further and better assess  the potential of folds  as particle reservoirs, the dynamics of this release should be characterized. 
We therefore plot in Fig. \ref{fig:dynamics} the temporal evolution of $A_{e}$ for all studied rafts.
The back compression does not  only limit the magnitude of the particle release but  also its duration. The latter is always less than $0.5s$, to be compared with $2s$ to $4s$ for similar compression achieved from the front side. Said differently, the incomplete character of the relaxation could, at first glance, be explained by a premature arrest of the process. Yet, a more detailed analysis of our data points toward  more complex phenomenon. 

Let us start considering the front compressed rafts, see Fig. \ref{fig:dynamics}(b). For interpreting the release dynamics, it may be useful to make an analogy with an electric system.  In this frame, the instantaneous rate of release,  or flow, $-\varphi_{e}\dot{A}_{e}$, corresponds to the current $i$. The flow is subjected to some resistance, $R$, which we assume to be constant. The confined area can be modeled as a capacitor of capacitance $C$, whose charge $q$ corresponds, at first order, to  the excess of particles. Thus,  for any instant $t$,  $q=\varphi_{j}(A_{r}-A_c)-\varphi_{e} A_{e}$, where  $\varphi_{j}(A_{r}-A_c)$ represents the initial excess of particles and $\varphi_{e} A_{e}$ the  escaped particles. Noting $U$ the potential difference, the  capacitor discharge trough a resistor $R$ is given by 
\begin{equation}
      C \frac{dU}{dt}+\frac{U}{R}=0\\
      \label{eq:capacitor}
\end{equation}
This equation can be solved for given initial conditions and provide $U(t)$, and by extension $i(t)=U(t)/R$.

In the present system, the first term of Eq. (\ref{eq:capacitor}), $C {dU}/{dt}={dq}/{dt}=i$, corresponds to  $-\varphi_{e}\dot{A}_{e}$ and the second one, ${U}/{R}={q}/{RC}$, to $[\varphi_{j}(A_{r}-A_c)-\varphi_{e} A_{e}]/RC$. The discharge equation therefore becomes
\begin{equation}
          \dot{A}_{e} + \frac{A_{e}-(A_r-A_c)\varphi_{j}/\varphi_{e}}{RC}=0
   \end{equation}
and the solution verifying the initial condition $A_e(t=0)=0$ is then:
\begin{equation}
          A_{e}=A_0(1-e^{-t/\tau})
          \label{eq:discharge}
   \end{equation}

with $A_0={(A_r-A_c)}{\varphi_{j}}/{\tau}{\varphi_{e}}$, the final escaped surface and $\tau=RC$, the typical release time constant.

To evaluate the relevance of our analogy, the experimental evolution of $A_e$ is fitted by Eq. (\ref{eq:discharge}) letting $A_0$ and $\tau$ be adjusted to minimize the sum of square residuals. The agreement is excellent, see Fig. \ref{fig:dynamics}(b). 
The biggest deviations are indeed observed either at the very beginning or at the end. The former is attributed to initial perturbations caused for example by capillary wave, by a possible short transient regime  or by the  difficulty to precisely identify the time origin.
The discrepancy found in the last instants may be caused by a premature arrest of the relaxation, which is discussed later.
The values of $\tau$  and $A_0/\tau$ produced by the fitting procedure are plotted in the inset as stars and squares, respectively. Both quantities seem to increase roughly linearly with $K$. Given the limited number of points, the interpretation remains tentative but one could postulate that $R$, the resistance to the flow, is  similar for all (front compressed) rafts, while $C$, the system capacitance, is directly proportional to the amount of particles stored in the folds, and thus to $K$. The initial discharge rate given by $A_0/\tau$ also  increases with $K$, but the linear character is less pronounced. If  confirmed, it would indicate that $\varphi_j\varphi_e$ varies as $K^2$. 
Future experiments could be used to probe this scaling,  the variations of $\varphi_e$ being  important in view of using folds as particles reservoir.


Let us now focus on  back compressed rafts. The experimental  points of Fig. \ref{fig:dynamics}(a) cannot be immediately identified with an exponential decay, which questions the previous interpretation. The latter considers the emptying of a reservoir subjected to a pressure, which  decreases linearly with $A_e$, and to  losses, which are  proportional to the flow, i.e. scaling as $\dot{A}_e$. Given the very limited decrease of $\Pi$ observed during the relaxation of back compressed rafts, it is legitimate to consider a constant pressure. Keeping the rest unchanged leads to   a simple linear increase of $A_e$ with $t$, which reads:
\begin{equation}
          A_{e}=\frac{\Pi_0}{\varphi_e R} t
          \label{eq:discharge_2}
   \end{equation}
Here $\Pi_0$ is the constant pressure applied at the back of the raft, $\varphi_e$ and $R$ are unchanged and represent the escaped particle density and the resistance to the flow, respectively. To test this model, the experimental results are fitted by linear functions, see dashed lines in Fig. \ref{fig:dynamics}(a). Note that  not zero intercepts are enabled since the experimental curves show an initial step. Beside this, the agreement is reasonable with the largest deviations observed at the end of the process.  Interestingly, assuming that $\Pi_0 \propto A_r-A_c$, we expect that the slope ${\Pi_0}/(\varphi_e R)$  of fixed size rafts (constant $A_r$) is proportional to $K$. This is indeed in good agreement with our data, see inset of Fig. \ref{fig:dynamics}(a), which shows - despite the limited number of points - a linear variation of the  fitted slopes with $K$.
These curves can therefore be seen as classical emptying of  granular silos. Indeed, the well known  Beverloo law predicts for given particle and orifice sizes, a constant flow rate \cite{beverloo1961flow}. The origin of this law remains controversial and neither the Janssen effect \cite{janssen1895versuche} nor the  free falling arch approach seem to provide the correct view \cite{rubio2015disentangling}. Furthermore, the dependency of the flow rate ($\dot{A}_e$) with the pressure  remains unclear. Some experiments evidence a total independence  \cite{aguirre2010pressure}, while others show  - under certain   circumstances -  a proportional relation \cite{peng2021external},  in agreement with our findings.
Yet, it is worth noting that these results, i.e.  $A_e(t) \propto t$ and  $\dot{A}_e \propto K$, are also compatible with the premature arrest of exponential release as found for front compressed rafts. One must here keep in mind that the limited character  of the  current data does allow to choose one of the two models. 

Whatever the  chosen function, the question of why and when  the relaxation process get arrested remains open. We attribute the ending to the existence of a yield stress, at least for back compressed rafts. Thus as long as the pressure $\Pi$ is above certain critical value, flow can occur. Small decrease of $\Pi$, even limited and at first order negligible (hypothesis of constant $\Pi_0$), can then stop the flow. Coming back to the electrical analogy, everything happens as if compressed rafts were granular diodes. When the diode is mounted in the appropriate direction, here corresponding to the front compression, the capacitor constituted by the particles stored in the folds can (quasi)-totally discharge, the current passing through a constant resistor $R$. Yet, if the compression occurs from the back, the diode,  mounted in the opposite direction, stops - after some leaking current has passed - the capacitor discharge.

In the context of compressed particle-laden interfaces, this diode effect can be understood in the light of granular framework, which describes the development of force chain network  during the compression.
By moving one barrier while keeping the other fixed, chains build up starting from the moving barrier and propagate via particle-particle contacts. For some of these contacts,  chains divide forming two or more  branches. Considering the monodisperse character of the particles, and by extension their hexagonal close packing, the orientation of these chains and branches  are expected to be  found in a cone, whose axis is the compression axis and angle is $60^{\circ}$. Consequently, for rafts compressed from the back side, the network  transmits the stress toward the front barrier but an important portion of it is redirected on each side of the orifice. The keystones, i.e. particles for which the chains branch, are in the back of the raft and therefore not easily removed. Arches may form that block the unjamming, which only occurs in a narrow corridor whose width is more or less the orifice width. In contrast, when  rafts are compressed from the front, the gate opening causes  the removal of  keystone particles found right behind it, leading the force chain  network to collapse, and therefore  triggering a quasi-total unjamming of the confined area.  A careful inspection of the unjammed  zone observed at the beginning of the process (see for example Fig. \ref{fig:2}, $t_{f,2}$) let see a conical shape whose angle is very close to $60^{\circ}$, in perfect agreement with the expected network structure. Finally, this interpretation also explains why  under certain circumstances, such as  very strong compressions, also front compressed rafts may give rise to a force chain structure that can,  at least partially, arrest the unjamming, see Fig. $K=67\%$ (circles) in Fig.  \ref{fig:dynamics}(b).



\section{Conclusions}
The different  behaviors observed during the relaxation of strongly compressed rafts evidence the importance of the compression direction on these systems. For back compressed rafts, the relaxation is incomplete while it is quasi-total for front compressed rafts. Practically, these results demonstrate that the processability of these interfaces highly depends on their history. In the present configuration, the  flowability  is primarily set by the compression direction, the level of compression playing a secondary role. This important finding should be accounted for in industrial processes. It is also of importance while considering the  self-healing properties of these interfaces. Indeed,  in this simple uni-axial geometry, the capacity of particles stored into large folds to migrate and stabilize uncovered areas is almost total for favorable compression direction, while it is strongly hindered in the opposite case. The dynamics of the particle release itself is strongly affected by the compression direction. While the current data remain limited, we observe similar trends in on-going tests performed using different raft size, orifice width and particle size.

These results are interpreted in the light of the framework developed for granular  matter, which demonstrates the importance of chain forces to transmit  stress. Thus, the direction of these chains, and more precisely their branching, is found to be essential as evidenced by triggering a local relaxation  along  the compression axis. For front compressed rafts, the local unjamming causes keystone particles to be removed leading to the network collapse and large scale unjamming. In contrast, for back compressed rafts the network redirects the stress laterally. As a result, only a limited number of particles found directly behind the orifice can flow, limiting the relaxation to a narrow corridor. 
To visualize these effects, the analogy with an electrical circuit can be made. The folds are like a capacitor and the particle flow like a current through a resistor. The force chain network can then be seen as a diode which let, or not, the current flow depending on its orientation.  

Finally, it is worth noting that our results and interpretation give rise to many open questions which need to be treated in  future research. From our point of view, future investigations should aim in understanding what influences the properties of the chain force network. In this frame, we can of course think of the raft and trough geometry, and especially of   the raft length, its width and the size of the orifice. A second important aspect is the influence of the individual particles, and more particularly,  the role of friction, shape, but also possible contact lubrication, which could for example cause network aging. 

\section*{Acknowledgments}
We would like to thank the Austrian Science Fund (FWF) for the financial support under Grant No. P33514-N. We also acknowledge  Graz University of Technology for its technical support, especially the Institute of Hydraulic Engineering and Water Resources Management for having borrowed us the high-speed camera and  the Institute of  Process and Particle Engineering for having helped us sieving the particles. C.P  acknowledges Elise Lorenceau and Anne-Laure Biance for fruitful discussions.

\section*{Appendix}

This Appendix presents 1) the preparation of the elastic barrier, 2) the method used to calibrate it and 3) the deformation equations, which are then used to deduce the lineic pressure acting on it.

\subsection*{1) Elastic rubber preparation}
The elastic rubber strings are produced in house by injecting a freshly prepared (1:1) mixture of Zhermack Elite Double 8 basis and catalyst (Zhermac Spa) into  glass capillaries. After the elastomer reticulation has been completed,  the glass capillaries are
manually  removed. Using capillaries of various diameter, we produce strings whose  diameter ranges between $0.5$ and $1$ $\mathrm{mm}$. Their typical length is in the range of $10$ $\mathrm{cm}$, which is sufficient to be installed in  $6$ $\mathrm{cm}$ wide trough.

\subsection*{2) Rubber calibration}

\begin{figure}[h]
\centering
\includegraphics[width=0.45\linewidth]{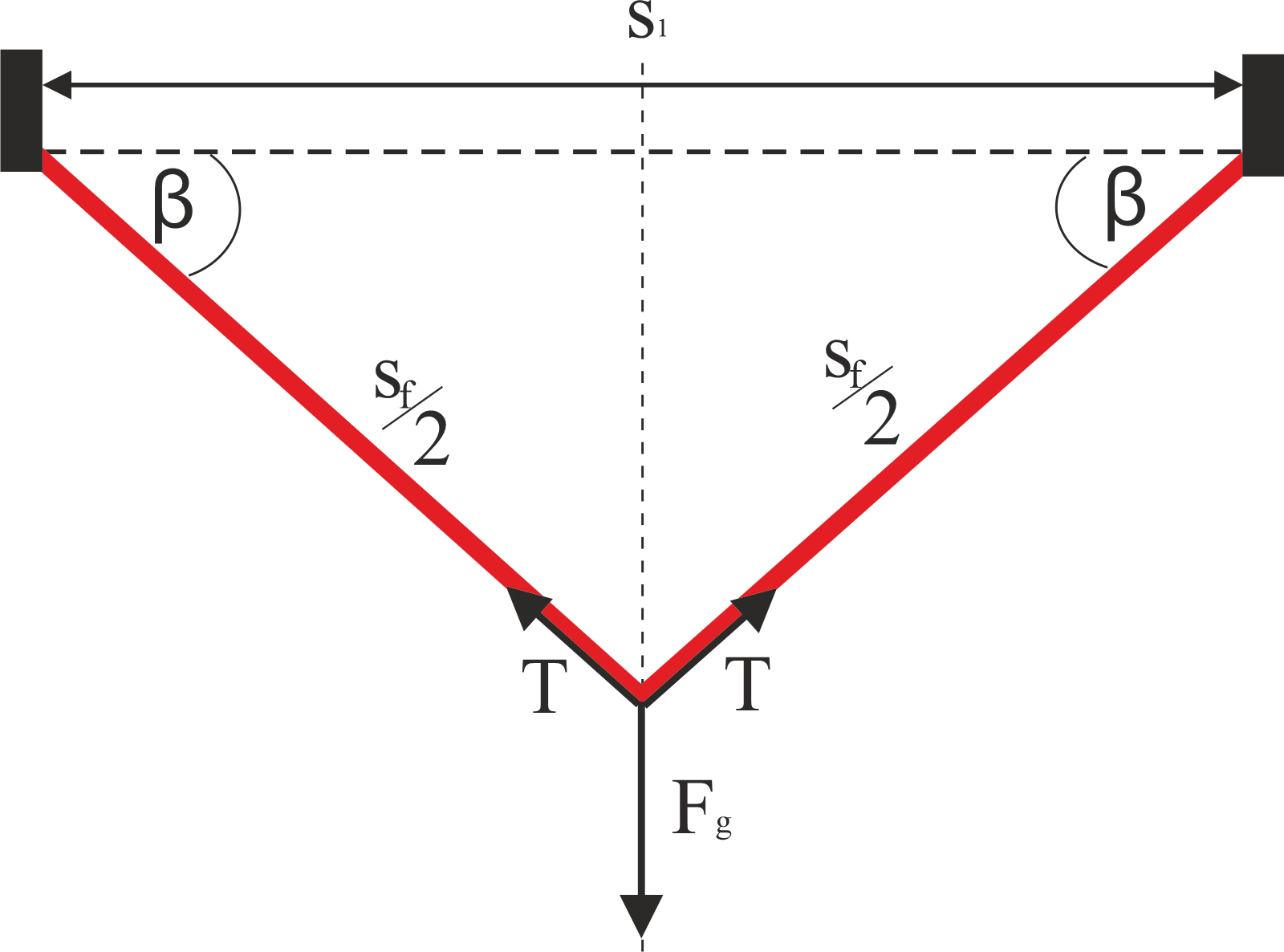}
\caption{Principle of the elastic calibration: deformation is measured as a function of the applied force.}
\label{fig:suppmat_1}
\end{figure}

The calibration of the rubber mostly consists in determining its  Young's modulus. The principle, detailed below, is simple, and consists in measuring the deflections produced by hanging known weights  to the rubber,  see Fig. \ref{fig:suppmat_1}. It was first presented in the PhD thesis of  \cite{PetitPHD2014}.

Consider an elastic string of relaxed length $s_0$ (not precisely known), attached between two holders separated by $s_1$  (black rectangles in Fig. \ref{fig:suppmat_1}). The elastic is further loaded using a known mass $m$ ($F_g=mg$) and reaches a stretched length $s_f$, which can easily be measured. The tensile stress, $\sigma$, is linearly proportional to the strain, $\epsilon$, and the proportionality coefficient, $E$, is the Young's modulus. This reads:
\begin{equation}
    \sigma =E \epsilon
    \label{eq:Hook1}
\end{equation}
The strain is equal to the relative extension and the tensile stress derives from  the tension $T$, providing:
\begin{equation}
    \epsilon=\frac{s_f-s_0}{s_0} \qquad \mathrm{and} \qquad \sigma=\frac{T}{\Sigma}
\end{equation}
with $\Sigma$, the cross-section area of the elastic string under tension. 
The Hook's law may now be rewritten as
\begin{equation}
    T=E ~\Sigma ~ \frac{s_f-s_0}{s_0}=E~ \Sigma_0~ \frac{s_f-s_0}{s_f}
    \label{eq:Hook2}
\end{equation}
where the variation of the cross-section area due to the stretching was introduced as
\begin{equation}
    \Sigma=\Sigma_0 \frac{s_0}{s_f}
\end{equation}
$\Sigma_0$ is the cross-section area of the non-deformed elastic string.
Under a small load, the force equilibrium may be written as: 
\begin{equation}
    F_g=2 T\sin{\beta} \qquad \mathrm{with} \qquad  \sin{\beta}  =\sqrt{1-\left(\frac{s_1}{s_f}\right)^2}
    \label{eq1}
\end{equation}
Here, $\beta$ is the angle between the rubber string and the horizontal.
Combining the above equations, one obtains the following relationship:
\begin{equation}
mg=2\lambda_0 \cdot \frac{s_f-s_0}{s_f}\sqrt{1-\left(\frac{s_1}{s_f}\right)^2}
\end{equation}
with the newly introduced parameter $\lambda_0$, defined by $\lambda_0=E \Sigma_0$.\\

\begin{figure}[h]
\centering
\includegraphics[width=0.45\linewidth]{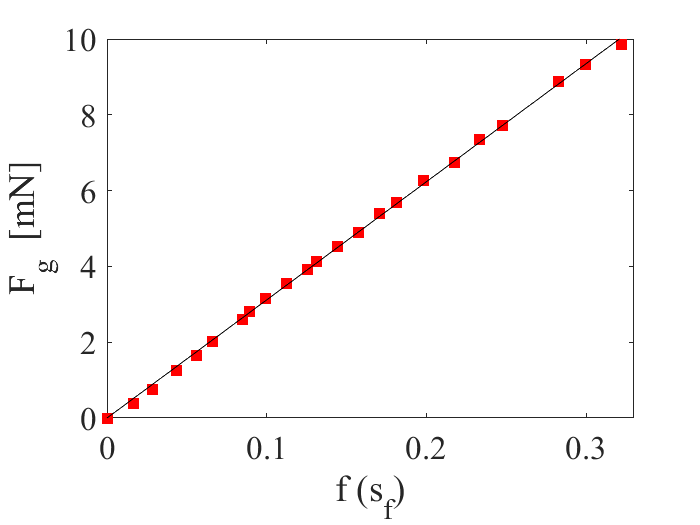}
\caption{Typical calibration curve and its linear fit providing $s_0$ (here $44.46mm$) and  $\lambda_0$ (here $0.3152N$). Function $f$ is defined as $f(s_f)=2\frac{s_f-s_0}{s_0}\sqrt{\left(1-\frac{s_1^2}{s_f^2}\right)}$. The coefficient of regression is $R^2=0.9995$. }
\label{fig:suppmat_2}
\end{figure}

Practically, the length of the deformed rubber $s_f$, is measured from pictures taken for different known masses $m$.  The undeformed rubber length, $s_0$, which is not precisely known and the parameter $\lambda_0$ are then obtained  by  finding the best fit to the measured data. This is done with the help of the \textit{fminsearch} function of Matlab. An illustrative curve is displayed in Fig. \ref{fig:suppmat_2}. The coefficient of regression is always very close to one, greater than 0.99. The 95\% confident intervals corresponds to an uncertainty of $\pm1\%$ on $\lambda_0$ and $s_0$. 
Knowing $\Sigma_0$, the cross section of the undeformed string, the modulus of elasticity is finally deduced as:
\begin{equation}
    E=\frac{\lambda_0}{\Sigma_0}
    \label{eq:E}
\end{equation}
Using the Elite Double 8, the measured $E$ is found to be in the expected  range of $0.1 MPa$. Note, that the two parameters $s_0$ and  $\lambda_0$ are also necessary to describe the deformation of the elastic string subjected to a constant lineic force, as  shown in the next section.

\subsection*{3) Deformation equations}

\begin{figure}[ht]
\centering
\includegraphics[width=0.45\linewidth]{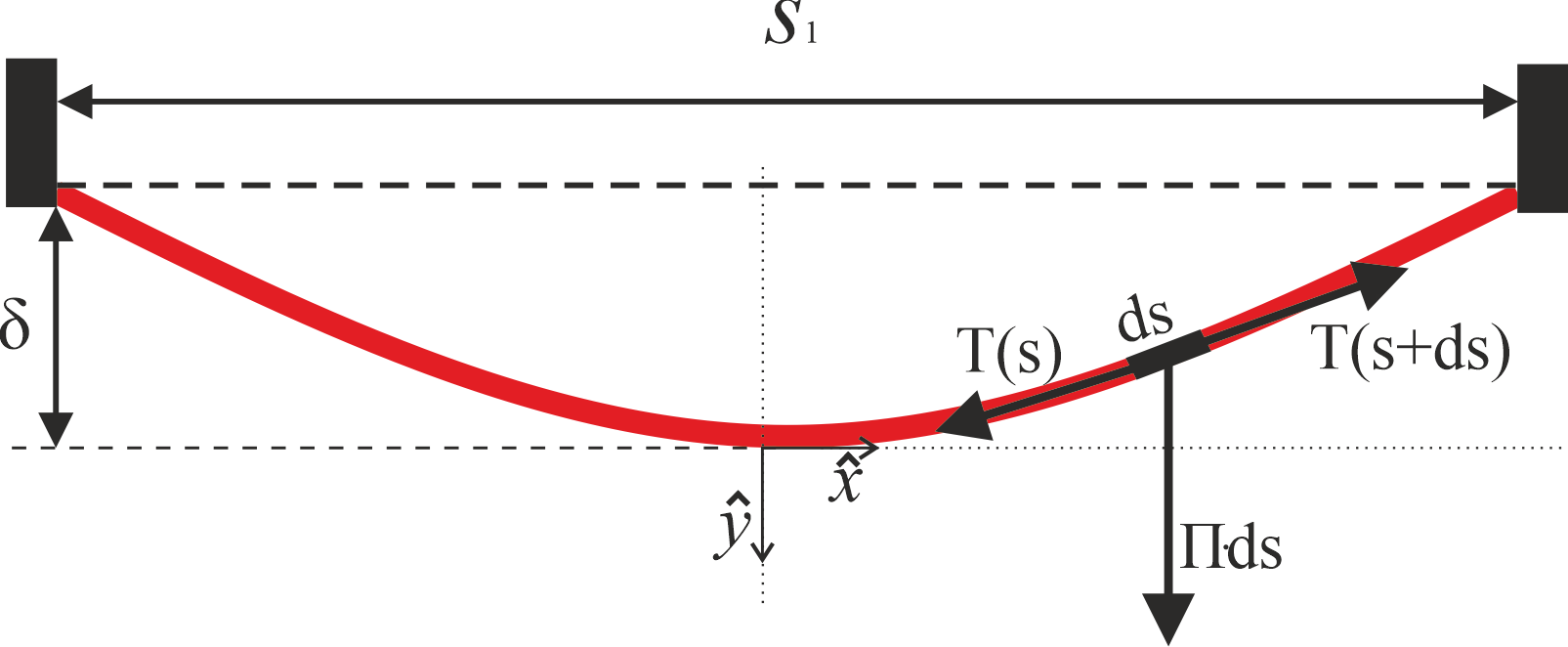}
\caption{Deformation of an elastic string due to the action of constant lineic  force $\Pi$.}
\label{fig:suppmat_3}
\end{figure}

The calibrated rubber string can then be used as a pressure sensor for the particle rafts in our experiments. To be quantitative,  the relationship between the elastic deflection, $\delta$, and the applied lineic pressure, $\Pi$, is needed. These quantities are defined in Fig. \ref{fig:suppmat_3}. The length of the stretched elastic is $s_f$, as previously defined. The position along the elastic can be described by the curvilinear ordinate  $s$ or by the cartesian coordinates $(x,y)$, the $x-$ and $y-$axes have unit vectors $\hat{x}$ and $\hat{y}$, respectively.

The following derivation is based on the well-known mathematical problem of a classical catenary \cite{catenary2015}. 
The tension $\vec{T}$ has a norm ($T$) and is collinear to the string  at any point $s \in [-s_f/2; s_f/2]$. Thus, naming $T_x$ and $T_y$ its horizontal and vertical components,  we get:

\begin{equation}
    \frac{\mathrm{d} y}{\mathrm{d} x}=\frac{T_y}{T_x}  \qquad \mathrm{and} \qquad  T=    \sqrt{{T_x}^2+{T_y}^2}
    \label{eq:tension8}
\end{equation}

Using the definition of the lineic pressure, $\Pi$, and applying the force balance to a string element of infinitesimal length $ds$ (see Fig. \ref{fig:suppmat_3}) provides:
\begin{equation}
    T_y=\int^{s_f/2}_{0} \Pi \mathrm{d}s = \Pi \int^{s_f/2}_{0}  \mathrm{d}s
    \label{eq:tension9}
\end{equation}
which leads with Eq. (\ref{eq:tension8}) to:
\begin{equation}
    \frac{\mathrm{d} y}{\mathrm{d} x}=\frac{\Pi}{T_x}\int^{s_f/2}_{0}  \mathrm{d}s=\frac{\Pi}{T_x}\int^{s_1/2}_{0} \frac{\mathrm{d}s}{\mathrm{d}x} \mathrm{d}x
    \label{eq:tension10}
\end{equation}
Eq. (\ref{eq:tension10}) can be differentiated with respect to $x$, leading to:
\begin{equation}
    \frac{\mathrm{d}^2 y}{\mathrm{d} x^2}=\frac{\Pi}{T_x} \frac{\mathrm{d}s}{\mathrm{d}x}
    \label{eq:tension11}
\end{equation}

To go further, the curvilinear ordinate is eliminated using:
\begin{equation}
    {\mathrm{d}s}^2= {\mathrm{d}x}^2+{\mathrm{d}y}^2={\mathrm{d}x}^2 \left(
    1+\left(\frac{\mathrm{d}y}{\mathrm{d}x}\right)^2 \right)
\end{equation}
which also writes:
\begin{equation}
\frac{\mathrm{d}s}{\mathrm{d}x}=\sqrt{
    1+\left(\frac{\mathrm{d}y}{\mathrm{d}x}\right)^2 }
\end{equation}
Eq. (\ref{eq:tension11}) becomes:
\begin{equation}
    \frac{\mathrm{d}^2 y}{\mathrm{d} x^2}=\frac{\Pi }{T_x}\sqrt{
    1+\left(\frac{\mathrm{d}y}{\mathrm{d}x}\right)^2 }
    \label{eq:tension4}
\end{equation}
It is  integrated with boundary conditions $\mathrm{d}y/\mathrm{d}x (x=0)=0$ and ${y(x=0)=0}$, providing:
\begin{equation}
    y(x)=\frac{T_x}{\Pi}\left( 
    \cosh{\left(\frac{\Pi}{T_x}x\right)}-1
    \right)
    \label{eq:15}
\end{equation}
Evaluating this expression for $x=s_1/2$ gives the central string deflection:
\begin{equation}
   \delta = \frac{T_x}{\Pi}\left( 
    \cosh{\left(\frac{\Pi s_1}{2T_x}\right)}-1
    \right)
    \label{eq:deflection}
\end{equation}

In practise, to use this expression, it is necessary to explicit $T_x$ as a function of the known parameters, i.e. $s_1$, $s_0$ and $\lambda_0$. Combining Eq. (\ref{eq:tension8}) and (\ref{eq:tension9}), we get:
\begin{equation}
    T=   \sqrt{{T_x}^2+\left(\Pi \int^{s_f/2}_0 \mathrm{d}s\right)^2}
    \label{eq:tension17}
\end{equation}
Using the stress-strain relation from the calibration section (Eq. (\ref{eq:Hook2})) and the definition of the parameter $\lambda_0$ given by Eq. (\ref{eq:E}), we  identify $T$ with:
\begin{equation}
    T=\lambda_0 \frac{s_f-s_0}{s_f}
    \label{eq:tension18}
\end{equation}
and obtain:
\begin{equation}
\lambda_0 \frac{s_f-s_0}{s_f}=\sqrt{{T_x}^2+\left(\Pi \int^{s_f/2}_0 \mathrm{d}s\right)^2}
\label{eq:19}
\end{equation}
We then  eliminate $s_f$  using:
\begin{equation}
\begin{split}
   \frac{s_f}{2}= \int^{s_f/2}_0 \mathrm{d}s =
    \int^{s_1/2}_0{ \sqrt{    1+\left(\frac{\mathrm{d}y}{\mathrm{d}x}\right)^2}\mathrm{d}x} \\ =    \frac{T_x}{\Pi}\sinh{\left(\frac{\Pi s_1}{2T_x}\right)}
    \end{split}
    \label{eq:integ-1}
\end{equation}
Eq. (\ref{eq:19}) and(\ref{eq:integ-1}) provide the condition
\begin{equation}
    \lambda_0\frac{\frac{T}{\Pi}\sinh{\left(\frac{\Pi s_1}{2T_x}\right)}-\frac{s_0}{2}}{\frac{T_x}{\Pi}\sinh{\left(\frac{\Pi s_1}{2T_x}\right)}} =  \sqrt{T^2_x \left(1+ \sinh^2\left(\frac{\Pi s_1}{2T}  \right)\right)}
\end{equation}
or equivalently 
\begin{equation}
    T_x \cosh\left(\frac{\Pi s_1}{2T_x}  \right)=
    \lambda_0 \left(
    1-\frac{s_0\Pi}{2T_x\sinh{\left(\frac{\Pi s_1}{2T_x}\right)}}
    \right)
    \label{eq:22}
\end{equation}
The relation between $\delta$, the central deflection  of the elastic string and $\Pi$, the applied lineic pressure is therefore given by the  two equations (\ref{eq:deflection}) and (\ref{eq:22}). Note that, using Eq.  (\ref{eq:15}) and (\ref{eq:22}), the entire shape of the elastic is indeed described as a function of known and measurable parameters, namely $s_1$, $s_0$ and $\lambda_0$.

For small deformations, Eq. (\ref{eq:deflection}) and (\ref{eq:22}) can be simplified by a third-order series expansion with the approximation
\begin{equation}
    \frac{\mathrm{d}y}{\mathrm{d}x}\approx \frac{s_1 \Pi}{T_x}
\end{equation}
which provides the following relation between the lineic pressure $\Pi$ and the (small) central deflection $\delta$:
\begin{equation}
    \Pi=\frac{8\lambda_0}{s_1}
    \left[ 
    \left(1-\frac{s_0}{s_1}\right)\frac{\delta}{s_1}+
    \left(\frac{4 s_0}{3 s_1}-1\right)\left(\frac{2 \delta}{s_1}\right)^3
    \right]
    \label{eq:Force}
\end{equation}

When applying the lineic pressure measurement method described above, a non-negligible uncertainty must be taken into account. 
The accuracy depends strongly on the difference $s_1-s_0$ as shown below:
\begin{equation}
  \frac{\Delta \Pi}{\Pi}=2\frac{\Delta s_1}{s_1}+
  \frac{s_1 \Delta s_0+s_0\Delta s_1}{s_1 (s_1-s_0)}
    \label{eq:error}
\end{equation}
In the present work, we evaluate this uncertainty to approximately 15\%.

\bibliographystyle{ieeetr}

\begin{thebibliography}{51}
\providecommand{\natexlab}[1]{#1}
\providecommand{\url}[1]{\texttt{#1}}
\expandafter\ifx\csname urlstyle\endcsname\relax
  \providecommand{\doi}[1]{doi: #1}\else
  \providecommand{\doi}{doi: \begingroup \urlstyle{rm}\Url}\fi

\bibitem[Ramsden(1903)]{Ramsden1903}
W.~Ramsden.
\newblock "separation of solids in the surface-layers of solutions and
  'suspensions' (observations on surface-membranes, bubbles, emulsions, and
  mechanical coagulation). preliminary account.".
\newblock \emph{Proceedings of the Royal Society of London}, 72\penalty0
  (479):\penalty0 156--164, August 1903.
\newblock \doi{10.1098/rspl.1903.0034}.

\bibitem[Pickering(1907)]{Pickering1907}
S.~U. Pickering.
\newblock Emulsions.
\newblock \emph{Journal of the Chemical Society}, 91, 1907.
\newblock \doi{10.1039/ct9079102001}.

\bibitem[Pitois and Rouyer(2019)]{Pitois_2019}
Olivier Pitois and Florence Rouyer.
\newblock Rheology of particulate rafts, films, and foams.
\newblock \emph{Current Opinion in Colloid \& Interface Science}, 43:\penalty0
  125--137, 2019.
\newblock ISSN 1359-0294.
\newblock \doi{https://doi.org/10.1016/j.cocis.2019.05.004}.
\newblock URL
  \url{https://www.sciencedirect.com/science/article/pii/S1359029418301365}.
\newblock Rheology.

\bibitem[Binks(2002{\natexlab{a}})]{Binks2002}
B.~P. Binks.
\newblock Particles as surfactants - similarities and differences.
\newblock \emph{Current Opinion In Colloid \& Interface Science}, 7\penalty0
  (1-2):\penalty0 21--41, March 2002{\natexlab{a}}.
\newblock \doi{10.1016/S1359-0294(02)00008-0}.

\bibitem[Herzig et~al.(2007)Herzig, White, Schofield, Poon, and
  Clegg]{Herzig_2007}
E.~M. Herzig, K.~A. White, A.~B. Schofield, W.~C.~K. Poon, and P.~S. Clegg.
\newblock Bicontinuous emulsions stabilized solely by colloidal particles.
\newblock \emph{Nature Materials}, 6:\penalty0 966--971, 2007.
\newblock \doi{10.1038/nmat2055}.
\newblock Rheology.

\bibitem[Cates and Clegg(2008)]{Cates_2008}
Michael~E Cates and Paul~S Clegg.
\newblock Bijels: a new class of soft materials.
\newblock \emph{Soft Matter}, 4\penalty0 (11):\penalty0 2132--2138, 2008.

\bibitem[Aussillous and Qu{\'e}r{\'e}(2001)]{Aussillous2001}
P.~Aussillous and D.~Qu{\'e}r{\'e}.
\newblock Liquid marbles.
\newblock \emph{Nature}, 411\penalty0 (6840):\penalty0 924--927, June 2001.
\newblock \doi{10.1038/35082026}.

\bibitem[Abkarian et~al.(2013)Abkarian, Proti{\`e}re, Aristoff, and
  Stone]{Abkarian_2013}
Manouk Abkarian, Suzie Proti{\`e}re, Jeffrey~M Aristoff, and Howard~A Stone.
\newblock Gravity-induced encapsulation of liquids by destabilization of
  granular rafts.
\newblock \emph{Nature communications}, 4\penalty0 (1):\penalty0 1--8, 2013.

\bibitem[Jambon-Puillet et~al.(2018)Jambon-Puillet, Josserand, and
  Proti\`ere]{Jambon_2018}
Etienne Jambon-Puillet, Christophe Josserand, and Suzie Proti\`ere.
\newblock Drops floating on granular rafts: a tool for liquid transport and
  delivery.
\newblock \emph{Langmuir}, 34\penalty0 (15):\penalty0 4437--4444, 2018.

\bibitem[Pike et~al.(2002)Pike, Richard, Foster, and Mahadevan]{Pike1211}
Nathan Pike, Denis Richard, William Foster, and L.~Mahadevan.
\newblock How aphids lose their marbles.
\newblock \emph{Proceedings of the Royal Society of London B: Biological
  Sciences}, 269\penalty0 (1497):\penalty0 1211--1215, 2002.
\newblock ISSN 0962-8452.
\newblock \doi{10.1098/rspb.2002.1999}.

\bibitem[Kralchevsky et~al.(2001)Kralchevsky, Denkov, and
  Danov]{Kralchevsky_2001}
Peter~A. Kralchevsky, Nikolai~D. Denkov, and Krassimir~D. Danov.
\newblock Particles with an undulated contact line at a fluid interface:
  Interaction between capillary quadrupoles and rheology of particulate
  monolayers.
\newblock \emph{Langmuir}, 17\penalty0 (24):\penalty0 7694--7705, 2001.
\newblock \doi{10.1021/la0109359}.
\newblock URL \url{http://dx.doi.org/10.1021/la0109359}.

\bibitem[Garbin(2019)]{Garbin_2019}
Valeria Garbin.
\newblock Collapse mechanisms and extreme deformation of particle-laden
  interfaces.
\newblock \emph{Current Opinion in Colloid \& Interface Science}, 39:\penalty0
  202--211, 2019.
\newblock ISSN 1359-0294.
\newblock \doi{https://doi.org/10.1016/j.cocis.2019.02.007}.
\newblock URL
  \url{https://www.sciencedirect.com/science/article/pii/S1359029418301122}.
\newblock Special Topic Section: Outstanding Young Researchers in Colloid and
  Interface Science.

\bibitem[Reynaert et~al.(2007)Reynaert, Moldenaers, and
  Vermant]{reynaert2007interfacial}
Sven Reynaert, Paula Moldenaers, and Jan Vermant.
\newblock Interfacial rheology of stable and weakly aggregated two-dimensional
  suspensions.
\newblock \emph{Physical Chemistry Chemical Physics}, 9\penalty0 (48):\penalty0
  6463--6475, 2007.

\bibitem[Cicuta et~al.(2003)Cicuta, Stancik, and Fuller]{Cicuta2003}
P.~Cicuta, E.~J. Stancik, and G.~G. Fuller.
\newblock Shearing or compressing a soft glass in 2{D}: Time-concentration
  superposition.
\newblock \emph{Physical Review Letters}, 90\penalty0 (23):\penalty0 236101,
  June 2003.
\newblock \doi{10.1103/PhysRevLett.90.236101}.

\bibitem[Lagubeau(2010)]{Lagubeau2010}
G.~Lagubeau.
\newblock PhD thesis, 2010.

\bibitem[Aveyard et~al.(2000)Aveyard, Clint, Nees, and Quirke]{Aveyard2000}
R.~Aveyard, J.~H. Clint, D.~Nees, and N.~Quirke.
\newblock Structure and collapse of particle monolayers under lateral pressure
  at the octane/aqueous surfactant solution interface.
\newblock \emph{Langmuir}, 16\penalty0 (23):\penalty0 8820--8828, November
  2000.
\newblock \doi{10.1021/la000060i}.

\bibitem[Monteux et~al.(2007)Monteux, Kirkwood, Xu, Jung, and
  Fuller]{monteux2007determining}
C{\'e}cile Monteux, John Kirkwood, Hui Xu, Eric Jung, and Gerald~G Fuller.
\newblock Determining the mechanical response of particle-laden fluid
  interfaces using surface pressure isotherms and bulk pressure measurements of
  droplets.
\newblock \emph{Physical chemistry chemical physics}, 9\penalty0 (48):\penalty0
  6344--6350, 2007.

\bibitem[Liu and Nagel(1998)]{liu1998jamming}
Andrea~J Liu and Sidney~R Nagel.
\newblock Jamming is not just cool any more.
\newblock \emph{Nature}, 396\penalty0 (6706):\penalty0 21--22, 1998.

\bibitem[Sollich et~al.(1997)Sollich, Lequeux, H{\'e}braud, and
  Cates]{sollich1997rheology}
Peter Sollich, Fran{\c{c}}ois Lequeux, Pascal H{\'e}braud, and Michael~E Cates.
\newblock Rheology of soft glassy materials.
\newblock \emph{Physical review letters}, 78\penalty0 (10):\penalty0 2020,
  1997.

\bibitem[H{\'e}braud and Lequeux(1998)]{hebraud1998mode}
Pascal H{\'e}braud and Fran{\c{c}}ois Lequeux.
\newblock Mode-coupling theory for the pasty rheology of soft glassy materials.
\newblock \emph{Physical review letters}, 81\penalty0 (14):\penalty0 2934,
  1998.

\bibitem[Beltramo et~al.(2017)Beltramo, Gupta, Alicke, Liascukiene, Gunes,
  Baroud, and Vermant]{beltramo2017arresting}
Peter~J Beltramo, Manish Gupta, Alexandra Alicke, Irma Liascukiene, Deniz~Z
  Gunes, Charles~N Baroud, and Jan Vermant.
\newblock Arresting dissolution by interfacial rheology design.
\newblock \emph{Proceedings of the National Academy of Sciences}, 114\penalty0
  (39):\penalty0 10373--10378, 2017.

\bibitem[Taccoen et~al.(2016)Taccoen, Lequeux, Gunes, and
  Baroud]{taccoen2016probing}
Nicolas Taccoen, Fran{\c{c}}ois Lequeux, Deniz~Z Gunes, and Charles~N Baroud.
\newblock Probing the mechanical strength of an armored bubble and its
  implication to particle-stabilized foams.
\newblock \emph{Physical Review X}, 6\penalty0 (1):\penalty0 011010, 2016.

\bibitem[Abkarian et~al.(2007)Abkarian, Subramaniam, Kim, Larsen, Yang, and
  Stone]{Abkarian2007}
M.~Abkarian, A.~B. Subramaniam, S.~H. Kim, R.~J. Larsen, S.~M. Yang, and H.~A.
  Stone.
\newblock Dissolution arrest and stability of particle-covered bubbles.
\newblock \emph{Physical Review Letters}, 99\penalty0 (18):\penalty0 188301,
  November 2007.
\newblock \doi{10.1103/PhysRevLett.99.188301}.

\bibitem[Timounay and Rouyer(2017)]{timounay2017viscosity}
Yousra Timounay and Florence Rouyer.
\newblock Viscosity of particulate soap films: approaching the jamming of 2d
  capillary suspensions.
\newblock \emph{Soft Matter}, 13\penalty0 (18):\penalty0 3449--3456, 2017.

\bibitem[Milner et~al.(1989)Milner, Joanny, and Pincus]{milner1989buckling}
Scott~Thomas Milner, JF~Joanny, and P~Pincus.
\newblock Buckling of langmuir monolayers.
\newblock \emph{EPL (Europhysics Letters)}, 9\penalty0 (5):\penalty0 495, 1989.

\bibitem[Helfrich(1973)]{helfrich1973elastic}
Wolfgang Helfrich.
\newblock Elastic properties of lipid bilayers: theory and possible
  experiments.
\newblock \emph{Zeitschrift f{\"u}r Naturforschung c}, 28\penalty0
  (11-12):\penalty0 693--703, 1973.

\bibitem[Cerda and Mahadevan(2003)]{Cerda2003}
E.~Cerda and L.~Mahadevan.
\newblock Geometry and physics of wrinkling.
\newblock \emph{Physical Review Letters}, 90\penalty0 (7):\penalty0 074302,
  February 2003.
\newblock \doi{10.1103/PhysRevLett.90.074302}.

\bibitem[Pocivavsek et~al.(2008)Pocivavsek, Dellsy, Kern, Johnson, Lin, Lee,
  and Cerda]{Pocivavsek2008a}
L.~Pocivavsek, R.~Dellsy, A.~Kern, S.~Johnson, B.~H. Lin, K.~Y.~C. Lee, and
  E.~Cerda.
\newblock Stress and fold localization in thin elastic membranes.
\newblock \emph{Science}, 320\penalty0 (5878):\penalty0 912--916, May 2008.
\newblock \doi{10.1126/science.1154069}.

\bibitem[Vella et~al.(2004)Vella, Aussillous, and Mahadevan]{Vella2004}
D.~Vella, P.~Aussillous, and L.~Mahadevan.
\newblock Elasticity of an interfacial particle raft.
\newblock \emph{Europhysics Letters}, 68\penalty0 (2):\penalty0 212--218,
  October 2004.
\newblock \doi{10.1209/epl/i2004-10202-x}.

\bibitem[Planchette et~al.(2012)Planchette, Lorenceau, and
  Biance]{Planchette2012a}
C.~Planchette, E.~Lorenceau, and A.~L. Biance.
\newblock Surface wave on a particle raft.
\newblock \emph{Soft Matter}, 8\penalty0 (8):\penalty0 2444--2451, 2012.
\newblock \doi{10.1039/c2sm06859a}.

\bibitem[Petit et~al.(2016)Petit, Biance, Lorenceau, and
  Planchette]{pre_bidisperse}
Pauline Petit, Anne-Laure Biance, Elise Lorenceau, and Carole Planchette.
\newblock Bending modulus of bidisperse particle rafts: Local and collective
  contributions.
\newblock \emph{Physical Review E}, 93\penalty0 (042802), 2016.

\bibitem[Cicuta and Vella(2009)]{Cicuta2009}
P.~Cicuta and D.~Vella.
\newblock Granular character of particle rafts.
\newblock \emph{Physical Review Letters}, 102\penalty0 (13):\penalty0 138302,
  April 2009.
\newblock \doi{10.1103/PhysRevLett.102.138302}.

\bibitem[Pitois et~al.(2015)Pitois, Buisson, and Chateau]{Pitois2015}
O.~Pitois, M.~Buisson, and X.~Chateau.
\newblock On the collapse pressure of armored bubbles and drops.
\newblock \emph{The European Physical Journal E}, 38\penalty0 (5):\penalty0 48,
  May 2015.
\newblock ISSN 1292-895X.
\newblock \doi{10.1140/epje/i2015-15048-9}.
\newblock URL \url{https://doi.org/10.1140/epje/i2015-15048-9}.

\bibitem[Vella et~al.(2006)Vella, Kim, Aussillous, and Mahadevan]{Vella2006}
D.~Vella, H.~Y. Kim, P.~Aussillous, and L.~Mahadevan.
\newblock Dynamics of surfactant-driven fracture of particle rafts.
\newblock \emph{Physical Review Letters}, 96\penalty0 (17):\penalty0 178301,
  May 2006.
\newblock \doi{10.1103/PhysRevLett.96.178301}.

\bibitem[Basavaraj et~al.(2006)Basavaraj, Fuller, Fransaer, and
  Vermant]{basavaraj2006packing}
MG~Basavaraj, GG~Fuller, Jan Fransaer, and Jan Vermant.
\newblock Packing, flipping, and buckling transitions in compressed monolayers
  of ellipsoidal latex particles.
\newblock \emph{Langmuir}, 22\penalty0 (15):\penalty0 6605--6612, 2006.

\bibitem[Saavedra et~al.(2018)Saavedra, Elettro, and
  Melo]{saavedra2018progressive}
Oscar Saavedra, Herv{\'e} Elettro, and Francisco Melo.
\newblock Progressive friction mobilization and enhanced janssen's screening in
  confined granular rafts.
\newblock \emph{Physical Review Materials}, 2\penalty0 (4):\penalty0 043603,
  2018.

\bibitem[Tordesillas et~al.(2011)Tordesillas, Lin, Zhang, Behringer, and
  Shi]{tordesillas2011}
Antoinette Tordesillas, Qun Lin, Jie Zhang, RP~Behringer, and Jingyu Shi.
\newblock Structural stability and jamming of self-organized cluster
  conformations in dense granular materials.
\newblock \emph{Journal of the Mechanics and Physics of Solids}, 59\penalty0
  (2):\penalty0 265--296, 2011.

\bibitem[Peters et~al.(2005)Peters, Muthuswamy, Wibowo, and
  Tordesillas]{peters2005}
JF~Peters, M~Muthuswamy, J~Wibowo, and A~Tordesillas.
\newblock Characterization of force chains in granular material.
\newblock \emph{Physical review E}, 72\penalty0 (4):\penalty0 041307, 2005.

\bibitem[Planchette et~al.(2018)Planchette, Lorenceau, and
  Biance]{planchette2018rupture}
Carole Planchette, Elise Lorenceau, and Anne-Laure Biance.
\newblock Rupture of granular rafts: effects of particle mobility and
  polydispersity.
\newblock \emph{Soft Matter}, 14\penalty0 (31):\penalty0 6419--6430, 2018.

\bibitem[Binks(2002{\natexlab{b}})]{binks_2002}
Bernard~P Binks.
\newblock Particles as surfactants—similarities and differences.
\newblock \emph{Current opinion in colloid \& interface science}, 7\penalty0
  (1-2):\penalty0 21--41, 2002{\natexlab{b}}.

\bibitem[Garbin(2013)]{garbin_2013}
Valeria Garbin.
\newblock Colloidal particles: Surfactants with a difference.
\newblock \emph{Phys. Today}, 66\penalty0 (10):\penalty0 68--69, 2013.

\bibitem[Arganda-Carreras et~al.(2017)Arganda-Carreras, Kaynig, Rueden,
  Eliceiri, Schindelin, Cardona, and Sebastian~Seung]{arganda2017trainable}
Ignacio Arganda-Carreras, Verena Kaynig, Curtis Rueden, Kevin~W Eliceiri,
  Johannes Schindelin, Albert Cardona, and H~Sebastian~Seung.
\newblock Trainable weka segmentation: a machine learning tool for microscopy
  pixel classification.
\newblock \emph{Bioinformatics}, 33\penalty0 (15):\penalty0 2424--2426, 2017.

\bibitem[Thielicke and Stamhuis(2014)]{thielicke2014pivlab}
William Thielicke and Eize Stamhuis.
\newblock Pivlab--towards user-friendly, affordable and accurate digital
  particle image velocimetry in matlab.
\newblock \emph{Journal of open research software}, 2\penalty0 (1), 2014.

\bibitem[Thielicke and Sonntag(2021)]{thielicke2021particle}
William Thielicke and Ren{\'e} Sonntag.
\newblock Particle image velocimetry for matlab: Accuracy and enhanced
  algorithms in pivlab.
\newblock \emph{Journal of Open Research Software}, 9\penalty0 (1), 2021.

\bibitem[Beverloo et~al.(1961)Beverloo, Leniger, and Van~de
  Velde]{beverloo1961flow}
Wim~A Beverloo, Hendrik~Antonie Leniger, and J~Van~de Velde.
\newblock The flow of granular solids through orifices.
\newblock \emph{Chemical engineering science}, 15\penalty0 (3-4):\penalty0
  260--269, 1961.

\bibitem[Janssen(1895)]{janssen1895versuche}
HA~Janssen.
\newblock Versuche uber getreidedruck in silozellen.
\newblock \emph{Z. Ver. Dtsch. Ing.}, 39\penalty0 (35):\penalty0 1045--1049,
  1895.

\bibitem[Rubio-Largo et~al.(2015)Rubio-Largo, Janda, Maza, Zuriguel, and
  Hidalgo]{rubio2015disentangling}
Sara~Mar{\'\i}a Rubio-Largo, Alvaro Janda, D~Maza, I~Zuriguel, and RC~Hidalgo.
\newblock Disentangling the free-fall arch paradox in silo discharge.
\newblock \emph{Physical review letters}, 114\penalty0 (23):\penalty0 238002,
  2015.

\bibitem[Aguirre et~al.(2010)Aguirre, Grande, Calvo, Pugnaloni, and
  G{\'e}minard]{aguirre2010pressure}
Maria~Alejandra Aguirre, Juan~G Grande, Adriana Calvo, Luis~A Pugnaloni, and
  J-C G{\'e}minard.
\newblock Pressure independence of granular flow through an aperture.
\newblock \emph{Physical review letters}, 104\penalty0 (23):\penalty0 238002,
  2010.

\bibitem[Peng et~al.(2021)Peng, Zhou, Zhou, Miao, Cheng, Jiang, and
  Hou]{peng2021external}
Zheng Peng, Jiangmeng Zhou, Jiahao Zhou, Yuan Miao, Liyu Cheng, Yimin Jiang,
  and Meiying Hou.
\newblock External pressure dependence of granular orifice flow: Transition to
  beverloo flow.
\newblock \emph{Physics of Fluids}, 33\penalty0 (4):\penalty0 043313, 2021.

\bibitem[Petit(2014)]{PetitPHD2014}
Pauline Petit.
\newblock \emph{D{\'e}formation des interfaces complexes : des architectures
  savonneuses aux mousses de particules}.
\newblock PhD thesis, University of Lyon, 2014.

\bibitem[Borggräfe et~al.(2015)Borggräfe, Heiligers, Ceriotti, and
  McInnes]{catenary2015}
Andreas Borggräfe, Jeannette Heiligers, Matteo Ceriotti, and Colin~R. McInnes.
\newblock Shape control of slack space reflectors using modulated solar
  pressure.
\newblock \emph{Proceedings of the Royal Society A}, 471\penalty0 (2179), 2015.

\end{thebibliography}

\end{document}